\documentclass[12pt]{article}

\usepackage{amssymb,amsmath,amsfonts,amsthm,geometry,ulem,graphicx,caption,color,setspace,sectsty,comment,footmisc,natbib,pdflscape,subfigure,array,booktabs,longtable,diagbox,datetime,subcaption,layouts,tikz,xcolor}
\usepackage[colorlinks=true,linkcolor=blue!50!black,citecolor=blue!50!black,urlcolor=blue!50!black]{hyperref}
\newtheorem{theorem}{Theorem}
\newtheorem{lemma}{Lemma}
\newtheorem{definition}{Definition}

\newtheorem{proposition}{Proposition}

\newtheorem{example}{Example}
\sectionfont{\fontsize{14}{13}\selectfont\centering}
\subsectionfont{\fontsize{12}{13}\selectfont\centering}
\subsubsectionfont{\fontsize{12}{12}\mdseries}

\title{On the Dominance of Truth-telling \\ in Gradual Mechanisms\thanks{We are especially grateful to Andrew Mackenzie, and we also thank In\'acio B\'o, Sambuddha Ghosh, Manshu Khanna, Qingmin Liu, Leeat Yariv, and anonymous referees for their helpful comments.}}
\author{Wenqian Wang\thanks{Southwestern University of Finance and Economics; wang.wenqian.cd@outlook.com} \and Zhiwen Zheng\thanks{Southwestern University of Finance and Economics; zheng.zhiwen.cn@gmail.com}}
\date{May, 2026}

\begin{document}

\onehalfspacing

\begin{titlepage}
\maketitle

\begin{abstract}

\noindent Recent literature highlights the advantages of implementing social rules via dynamic game forms. We characterize when truth-telling remains a dominant strategy in gradual mechanisms, where agents progressively reveal their private information while acquiring information about others. Our first characterization hinges on the incentive-preservation of a basic transformation on gradual mechanisms called illuminating that transmits more information to an agent by partitioning her information sets. The second characterization is a single reaction-proofness condition where reaction refers to how one agent acts immediately after acquiring information about others. We apply these results to the second-price auction and the hierarchical exchange rule.

  \vspace{0.15in}

  \noindent\textbf{Keywords:} strategy-proofness, gradual mechanism, transformations on game forms, reaction-proofness, second-price auction, hierarchical exchange rule
  \end{abstract}
\setcounter{page}{0}
\thispagestyle{empty}
\end{titlepage}

\pagebreak \newpage

\onehalfspacing

\section{Introduction}
\label{sec:Intro}

In incomplete information environments, social planners face the challenge of eliciting private information from agents to determine social outcomes. While implementing strategy-proof social rules via their direct mechanisms---where agents simultaneously report all private information---has been standard practice since \cite{GB1973} and \cite{myerson1981}, a growing literature has begun to recognize that dynamic procedures may offer advantages in terms of strategic simplicity \citep{Li2017,pycia2023theory, bo2024pick,nagel2025measure,CW2024}, privacy protection \citep{liu2020preserving,haupt2024}, and credibility or transparency \citep{Akbarpour2020,hakimovms,moller2024transparent}.

Dynamic mechanisms naturally prompt the question of when they maintain (dominant strategy) incentive compatibility. Consider two bidders competing for a single item. In the standard sealed-bid second-price auction, truthful bidding is a dominant strategy. However, this property may not survive dynamic variants of the auction. Suppose instead that bidder $2$ observes bidder $1$'s bid before submitting her own. Then incentive compatibility breaks down: if bidder $2$ responds by bidding some $b_2 > b_1$ whenever bidder $1$ bids a pre-specified value $b_1$, and bids $0$ otherwise, then truthful bidding is no longer optimal for bidder $1$ when her private value for the item equals $b_1$.

Indeed, the potential advantages of dynamic mechanisms originate from their dynamic information flow, which inevitably enables agents to condition their actions on the information they acquire in the process. Therefore, the broadened strategy spaces may undermine the incentive compatibility of the social rules implemented by these dynamic mechanisms. In this paper, we address this question for strategy-proof social choice functions,\footnote{Strategy-proofness of a social choice function is equivalent to the (weak) dominance of truth-telling in its direct mechanism, which is considered the golden criterion of incentive compatibility.} which are widely applied in practice, by characterizing the dominance of truth-telling in their \textit{gradual mechanisms}. Our characterization helps a designer to pursue the advantages of dynamic mechanisms while preserving dominant strategy implementation.

Generally, a gradual mechanism can be regarded as the following information revelation procedure involving an administrator.\footnote{Call a dynamic mechanism an information revelation procedure if the agents report their private information straightforwardly by choosing from subsets of their private types at each decision node. Truth-telling is naturally defined in these procedures as a strategy where an agent consistently selects the subset containing her private type.}

\begin{quote}
    
  The administrator privately sends an information-gathering form and a personalized message to each active agent at each stage. The form contains a list of disjoint categories of the agent's possible private types, refining her previous report. The message conveys some information about the categories selected by other agents in their prior forms, based on which the agent checks her current form. The administrator keeps sending these forms and messages until she collects enough information to determine a social outcome.
\end{quote}

The above description reveals that a gradual mechanism arises from any information revelation procedure where (i) the truth-telling option is available and unique at any decision node, provided an agent has been truthful up to that point, and (ii) agents cannot provide contradictory information about themselves, even after deviating from truth-telling. Both restrictions are natural for an information revelation procedure to implement a social choice function. Further discussion about our chosen domain can be found in the concluding section.

The paper is structured as follows: Section \ref{secframe} introduces the mechanism design environment and the gradual mechanisms. Section \ref{secchara} presents our characterizations. Applications are presented in Section \ref{secapp}. Section \ref{seccon} concludes. Appendices collect omitted definitions and proofs. In the rest of the introduction, we provide an outline of our characterizations, a comparison with \cite{mackenzie2022}, a sketch of our applications, and a review of related literature.

\vspace{5mm}

\noindent\textit{Outline of characterizations.}
In direct mechanisms, every agent acts only once and cannot condition her choice on the actions of others, making truth-telling dominant as long as the social choice function is strategy-proof. We start our characterization by analyzing how gradual mechanisms lose incentive compatibility---truth-telling being (weakly) dominant for each agent---as they sequentialize the decision-making processes. This approach leads us to develop three basic transformations---splitting (SPL), coalescing (COA), and illuminating (ILL)---which, if applied in a proper sequence (where ILL is applied inversely), can transform any gradual mechanism into the direct mechanism implementing the same social choice function.\footnote{There is a strand of literature on the basic transformations on game forms that preserve certain equivalence relations or strategic features, see \cite{Thompson1952}, \cite{Kohlberg1986}, \cite{Bonannot1992}, \cite{Elmes1994}, \cite{Battigalli2020}, and \cite{Wang2024}. We relate our three basic transformations to this literature when they are introduced in Section 3, where some visual representations of the transformations are also included.} An SPL requires an agent to reveal some additional information to the administrator at terminal histories. A COA moves the available actions at an information set of an agent to the immediately preceding one, provided that the agent acquires no further information upon reaching the later information set. Naturally, neither of these two transformations changes the incentive compatibility of two gradual mechanisms linked by them. This observation leaves ILLs the only possible destroyers of incentive compatibility.

An ILL provides an agent, say agent $i$, with more information by dividing an information set of agent $i$ into two smaller ones, enabling her to choose different actions at these two newly created information sets based on the additional information acquired. For the gradual mechanism delivered by such an ILL to be incentive compatible, truth-telling for any other agent $j$ must remain a dominant strategy after taking agent $i$'s increased strategic flexibility into account, giving rise to a condition named \textit{incentive-preservation}. This delivers our first characterization: a gradual mechanism implementing a strategy-proof social choice function is incentive compatible if and only if all the ILLs are incentive-preserving in a chain of SPLs, COAs, and inverse ILLs that transforms it into the direct mechanism. As an application of this characterization, we identify a condition on a strategy-proof social choice function under which essentially only the direct mechanism for it can be incentive compatible.

Inspired by the above transformations-based analysis, we introduce a condition on gradual mechanisms, which we call \textit{reaction-proofness}, adapting a term from \cite{mackenzie2022}, that is equivalent to incentive compatibility in our environment. Like the incentive-preservation condition for ILLs, reaction-proofness only examines how an agent can react to others by reporting differently immediately after acquiring some information, i.e., at a pair of information sets following the same action at a common immediate predecessor, and requires that this type of reaction does not harm any other agent's incentives for truth-telling.

Both reaction-proofness and incentive-preservation require truth-telling to deliver weakly better outcomes for one agent even when some other agent can report differently depending on whether she is truthful or misrepresenting, thus sharing the spirit of \citeauthor{Li2017}'s (\citeyear{Li2017}) concept of obvious dominance. Drawing upon the idea of reducing the cognitive burden of contingent reasoning, which motivates Li's definition, we provide a stronger version of reaction-proofness, called \textit{strong reaction-proofness}. This condition bridges our paper and \cite{mackenzie2022} with which we now compare.

\vspace{5mm}
\noindent\textit{Comparison with \cite{mackenzie2022}.} Focusing on environments with private values and no consumption externalities, \citeauthor{mackenzie2022} study a class of dynamic mechanisms called \textit{menu mechanisms}, in which agents report their most preferred \textit{assignments} in a sequence of menus. For this type of information revelation, a truth-telling option always exists, even after one has deviated from truth-telling. Notice that this feature enables the authors to study \textit{everywhere dominance}, a more robust version of weak dominance that covers off-path histories. When preferences are \textit{strict}, the truth-telling option is unique. Furthermore, by assuming \textit{non-repeating} menus, an agent cannot later contradict an earlier report. Therefore, in environments with strict preferences, these non-repeating menu mechanisms give rise to a particular subclass of gradual mechanisms. Their reaction-proofness for the menu mechanisms---part of a sufficient condition for truth-telling to be everywhere dominant---requires that one agent's reaction does not harm another agent's truthful incentives because the latter agent's assignment has already been determined.

In contrast, we first allow consumption externalities and weak preferences. Second, we study gradual mechanisms that permit more flexible reports of private types, while focusing on weak dominance without restrictions on off-path behavior. Third, we provide two necessary-and-sufficient conditions of incentive compatibility in gradual mechanisms. Finally, in our strong reaction-proofness condition, we use indifference over outcomes rather than identical assignments for a sufficient condition of incentive compatibility in gradual mechanisms.

\vspace{5mm}
\noindent\textit{Sketch of applications.}
We first apply our results to a setting where an indivisible item is allocated to one of several bidders under the second-price auction rule with discrete private values and randomized tie-breaking.\footnote{A tie, i.e., multiple bidders submitting the same highest bids, arises with positive probability under discrete bidding levels. Fair randomization is a natural tie-breaking rule and is used in both theoretical analyses \citep[e.g.,][]{rothkopf1994role} and experiments \citep[e.g.,][]{kagel1987}.} The auctioneer decides to implement this rule in an ascending-price mechanism for strategic simplicity or privacy protection.\footnote{Compared to sealed-bid second-price auctions, ascending-price auctions elicit less information about the winning bidder's private value for the auctioned item \citep{ausubel2004}.} In this environment, the auctioneer faces a problem of information design. On one hand, keeping bidders' actions confidential may be costly in certain settings such as a physical auction house, and raise suspicions about the auctioneer's integrity. On the other hand, as illustrated by our opening example, an ascending-price auction where all bids are open may induce anxiety over bidding wars, and thereby fail to be incentive compatible. Therefore, it would be informative for the auctioneer to know the boundary of information transmission that maintains the incentive for truthful bidding.

The strong reaction-proofness condition underpins the well-received wisdom that any information from past price levels can be made public without undermining incentive compatibility \citep[e.g.,][]{krishna2009auction}. Regarding information transmission within a given price level, we show that all actions taken at the current price level $p$ can be made public once two bidders have chosen to stay at $p$, a feature we call \textit{open-by-two}. Finally, we show that this open-by-two ascending-price auction is \textit{transparent} in the sense that any further information transmission will destroy the incentive compatibility of truthful bidding, as no potential ILL is incentive-preserving. 


The second application concerns the hierarchical exchange rule proposed by \cite{papai2000strategyproof}, which generalizes the top trading cycles (TTC) algorithm \citep{shapley1974cores} for allocating indivisible objects to agents. \cite{CW2024} propose a class of gradual mechanisms, called renunciation-designation-assertion (RDA) mechanisms, that implement the TTC rule while easing agents' burden of contingent reasoning, extending the dynamic mechanisms in \cite{Troyan2019} and \cite{mandal2022obviously} to environments where obviously dominant strategy implementations are unavailable. We generalize RDA mechanisms to implement the hierarchical exchange rule and show their incentive compatibility through strong reaction-proofness.

Some extant dynamic mechanisms \citep[e.g.,][]{mackenzie2022,bo2024pick,nagel2025measure} mimic the TTC algorithm, iteratively asking agents to report their most preferred remaining object. In contrast, in the three sub-stages of RDA mechanisms, agents make different types of reports: an active agent (who must own a menu of objects) reveals whether her menu contains her top object in the renunciation sub-stage, which menu held by another owner contains her top object in the designation sub-stage, and her top object itself only in the final assertion sub-stage. Diverse reports at different sub-stages make incentive compatibility less closely tied to the strategy-proofness of the TTC algorithm. 

\vspace{5mm}
\noindent\textit{Other related literature.} Our paper is prompted by the literature initiated by \cite{Li2017} on obvious dominance and related criteria of strategic simplicity of dynamic mechanisms \citep{bade2017gibbard,Mackenzie2020,pycia2023theory,nagel2025measure,CW2024}. Rather than addressing obvious dominance or any particular refinement of weak dominance, we study a logically prior question: when and how can the direct mechanism of a strategy-proof social choice function be sequentialized without losing the weak dominance of truth-telling?

The aforementioned papers deliver several canonical dynamic mechanisms, which are special cases of gradual mechanisms or can be seen as such after relabeling each action as the set of types that choose it in equilibrium. Prominent examples include (i) \textit{gradual revelation mechanisms} \citep{bade2017gibbard}, (ii) \textit{round table mechanisms} \citep{Mackenzie2020}, (iii) \textit{millipede mechanisms} \citep{pycia2023theory}, and (iv) \textit{essential frames} \citep{nagel2025measure}.\footnote{Some features of these mechanisms fall outside our framework: (i) and (ii) relax perfect recall, while (ii), (iii), and (iv) allow chance moves.} Gradual mechanisms as defined in this paper have been used by \cite{CW2024} to study a simplicity measure inspired by the sure-thing principle \citep{savage1972}.

Beyond weak dominance, a related line of work studies dynamic implementation under ex-post incentive compatibility \citep{bergemann2005robust}.\footnote{Ex-post incentive compatibility requires truthful reporting to be a Nash equilibrium for every realized type profile and thus ensures robustness to agents' higher-order beliefs. Fix a social choice function, a dominant strategy dynamic implementation may need exponentially more communication than an ex-post dynamic implementation \citep{rubinstein2021exponential}.} \citet{nagel2024if} introduce \textit{as-if dominance}, which requires strategies in an ex-post equilibrium to be best responses against only ``naive" deviations by other agents. \cite{bo2024pick} characterize perfect ex-post incentive compatibility \citep{ausubel2004} in pick-an-object mechanisms, a subclass of menu mechanisms that guarantees each agent her last chosen assignment.\footnote{Similar to \cite{mackenzie2022}, they take off-path behavior into consideration.}

\section{Framework}
\label{secframe}

We outline the theoretical framework in this section. In the first two subsections, we introduce our mechanism design environment and the gradual mechanisms. In the last subsection, we provide some basic observations about the dominance of truth-telling in gradual mechanisms, which pave the way for our characterizations in the next section. 

\subsection{Environments}

A finite set $N$ of agents are interested in which social outcome in a finite set $X$ obtains. Each agent $i\in N$ has a (private) type space $\Theta_i$ in which a type $\theta_i$ corresponds to a complete and transitive preference ordering $R(\theta_i)$ over $X$. Let $P(\theta_i)$ denote the strict part of $R(\theta_i)$, i.e., for any $x,x' \in X$, $x P(\theta_i) x'$ if and only if $x R(\theta_i) x'$ and not $x' R(\theta_i) x$. Conventionally, a type profile is written as $\theta \in \Theta = \prod_{i\in N} \Theta_i$. For a given $\theta\in \Theta$, we use $\theta_i$ to denote the type of agent $i$ in $\theta$. We will also use the notation $\theta_{-i}$ (and $\theta_{-i,j}$) for a type profile of agents other than $i$ (other than $i$ and $j$). The social planner wishes to implement a social choice function (SCF) $f:\Theta\rightarrow X$ that assigns an outcome to a type profile. An SCF $f$ is strategy-proof if $f(\theta_i, \theta_{-i}) R(\theta_i) f(\theta_i', \theta_{-i})$ for all agents $i\in N$, all $\theta_i, \theta_i'\in \Theta_i$, and all $\theta_{-i}\in \Theta_{-i}$. 

We focus on implementation problems in which the main challenge for the social planner arises from private information, i.e., the private type of every agent is known only to herself. From \cite{GB1973}, we learn that a social planner could simply adopt the direct mechanism of a strategy-proof SCF, in which truthfully reporting one's private type is a (weakly) dominant strategy. In this paper, the social planner looks beyond direct mechanisms to harness the potential benefits of dynamic mechanisms. That is to say, the social planner considers every dynamic game form $G$ (with perfect recall and finite depth)---which can be described by the available actions $A_i$ of each player $i\in N$, the collection of (non-terminal and terminal) histories $\bar{H} = H\cup Z$ modeled by sequences of action profiles, information sets $\boldsymbol{H}_i$ for each player $i \in N$, and the outcome function $\mathcal{X}$ mapping terminal histories to outcomes in $X$---as a mechanism to implement an SCF $f$ by the equilibrium strategies in the incomplete information game $(G, \Theta)$.\footnote{We restrict to finite game trees as agents and outcomes are finite. We will formally adopt the framework of \cite{osborne1994course} and \cite{Battigalli2020}. For completeness, the definition of dynamic game forms is included in the appendix.}

\subsection{Gradual Mechanisms}

A gradual mechanism is a specific type of dynamic game form. Consider the following definition of a gradual mechanism $G$ implementing an SCF $f$ (in its truth-telling strategies).

\begin{definition}
    \label{def:gm}

    A gradual mechanism $G$ implementing an SCF $f$  is a dynamic game form $(H\cup Z, \{A_i, \boldsymbol{H}_i\}_{i\in N}, \mathcal{X})$ in which:

    \begin{enumerate}
        \item agents transmit information about their private types, i.e., for any agent $i \in N$, available actions at any decision node are non-empty subsets of $\Theta_i$;
        \item information being transmitted is gradually refined, i.e., for any agent $i\in N$ and any decision node $h\in H_i$ of agent $i$, we have (i) $a_i\cap a_i'=\varnothing$ for any two different available actions $a_i, a_i'\in A_i(h)$ and (ii) $\bigcup A_i(h) = \Theta_i(h)$, where $\Theta_i(h)$ is the last action of agent $i$ before $h$ (let $\Theta_i(h) = \Theta_i$ if agent $i$ has not acted);
        \item the outcomes are assigned according to the accrued information at the terminal histories, i.e., $\mathcal{X}(z) = f(\theta)$ for any $z\in Z$ and any $\theta\in \Theta(z) = \prod_{i\in N} \Theta_i(z)$.
    \end{enumerate}

\end{definition}

The direct mechanism of any SCF is also a gradual mechanism in which all agents simultaneously report their types at the initial history. Extending the notations in the above definition, for any history $h$, let $\Theta(h) = \prod_{i\in N} \Theta_i(h)$ denote the possible type profiles revealed by agents up to $h$. For any information set $\boldsymbol{h}_i$, let $\Theta_i(\boldsymbol{h}_i) = \bigcup_{h\in \boldsymbol{h}_i}\Theta_i(h)$ (which equals $\Theta_i(h)$ for any $h\in \boldsymbol{h}_i$ due to the perfect recall assumption) and let $\Theta_{-i}(\boldsymbol{h}_i) = \bigcup_{h\in \boldsymbol{h}_i}\Theta_{-i}(h)$ where $\Theta_{-i}(h) = \prod_{j\in N, j\neq i} \Theta_j(h)$. $\Theta_i(\boldsymbol{h}_i)$ and $\Theta_{-i}(\boldsymbol{h}_i)$ capture the information provided and acquired by agent $i$ at the information set $\boldsymbol{h}_i$, respectively. Let $\preceq$ denote the precedence relation over histories with $\prec$ being its asymmetric part. Note that $\Theta(\bar{h}) \subsetneq \Theta(h)$ if $h \prec \bar{h}$.

In dynamic game forms, a pure strategy $s_i \in S_i$ for any agent $i\in N$ chooses an action $s_i(h)$ at each decision node $h\in H_i$. In a gradual mechanism, we say a pure strategy $s_{i}$ for agent $i$ is unconditional for type $\theta_i$ if $\theta_i\in s_i(h)$ as long as $\theta_i\in\Theta_i(h)$ for all $h\in H_i$. As agents gradually refine their reports in gradual mechanisms, an unconditional strategy exists for every type of every agent. Though the unconditional strategies for a specific type are not unique, they all choose the unique truth-telling option at each decision node on the truthful paths. In a nutshell, they are behaviorally equivalent.


A complete profile $s \in S= \prod_{i \in N}S_i$ of pure strategies determines a unique terminal history $z(s)$. To simplify notation, we let $\mathcal{X}(s)$ denote $\mathcal{X}(z(s))$. We define the incentive compatibility of gradual mechanisms to be the (weak) dominance of truth-telling for each agent.

\begin{definition}
  \label{defi:ic}
  A gradual mechanism $G$ implementing an SCF $f$ is incentive compatible if $$\mathcal{X}(s_{\theta_i}, s_{-i}) R(\theta_i) \mathcal{X}(s_i, s_{-i})$$ for any agent $i\in N$, any type $\theta_i\in \Theta_i$, any unconditional strategy $s_{\theta_i}\in S_i$ for $\theta_i$, any strategy $s_i\in S_i$, and any strategy profile $s_{-i}\in S_{-i}=\prod_{j \in N, j\neq i}S_j$.
\end{definition}

\subsection{Preliminary Results}

We provide two propositions that will be used throughout without mentioning, whose proofs are in Appendix \ref{proofsec2}. Since an unconditional strategy exists for each private type of each agent in any gradual mechanism, we have the following proposition.

\begin{proposition}
  A gradual mechanism implementing an SCF $f$ is incentive compatible only if $f$ is strategy-proof.
  \label{propnec}
\end{proposition}

For a non-empty subset of agents $M \subseteq N$ and a profile $s_M \in S_M=\prod_{i \in M}S_i$ of pure strategies, we say a history $h$ is consistent with $s_M$ (or $s_M$ is consistent with $h$) if there exists $s' \in S$ with $s'_M=s_M$ such that $h$ is on the path determined by $s'$. In gradual mechanisms, we say a type profile $\theta$ is consistent with a strategy profile $s_M$ if the terminal history $z$ such that $\theta\in \Theta(z)$ is consistent with $s_M$. Since the outcomes are assigned according to the accrued information in gradual mechanisms, incentive compatibility is equivalent to the following.

\begin{proposition}
  \label{Unconditional2}
  A gradual mechanism $G$ implementing an SCF $f$ is incentive compatible if and only if for any agent $i\in N$ and any $\theta^1, \theta^2\in \Theta$ consistent with a common $s_{-i} \in S_{-i}$, it is the case that $f(\theta^1) R(\theta^1_i) f(\theta^2)$.
\end{proposition}

In this proposition, we use the values of the SCF $f$ directly, substituting the corresponding outcomes on terminal histories. This language will be adopted in our characterizations in the next section.

\section{Characterizations}
\label{secchara}

We offer two characterizations of incentive compatibility in gradual mechanisms. In the first subsection, we will develop a set of basic transformations on gradual mechanisms and characterize incentive compatibility based on these transformations. Next, we distill the intuition from this fine-grained analysis into a second characterization: a single condition named reaction-proofness.

To illustrate, we use the following simple voting scheme as a running example throughout this section. There are two voters $1$ and $2$ and three candidates $L$, $M$, and $R$. Slightly abusing notation, a type $L$ voter prefers $L$ over $M$ and further over $R$; a type $M$ voter prefers $M$ over $L$ and $R$ (inconsequentially, assume she is indifferent between $L$ and $R$); a type $R$ voter prefers $R$ over $M$ and further over $L$. The strategy-proof voting scheme elects candidate $L$ (or $R$) if both voters are of type $L$ ($R$, respectively) and candidate $M$ otherwise.\footnote{This SCF is known as a generalized median voter scheme \citep{barbera1993generalized}.}

\subsection{A Characterization Based on Basic Transformations}

We organize the basic transformations so as to reduce any gradual mechanism implementing any SCF into the direct mechanism implementing the same SCF, while also identifying a specific one --- namely ILL --- that changes the incentive compatibility of two gradual mechanisms linked by it. In the main text, we describe how these three basic transformations essentially modify the structure of a gradual mechanism and illustrate them with our running example. See Appendix \ref{sec:app_transformations} for their technical definitions.  

\subsubsection{Splitting}
Let $\boldsymbol{h}_i$ be an information set, and $\bar{\boldsymbol{h}}_i$ be the subset of terminal histories that pass through $\boldsymbol{h}_i$, in which agent $i$ chooses $a_i\in A(\boldsymbol{h}_i)$ at $\boldsymbol{h}_i$ and takes no further action thereafter. Let $a_i^1$ and $a_i^2$ be two non-empty disjoint subsets of $\Theta_i$ whose union is $a_i$. When $\bar{\boldsymbol{h}}_i$ is non-empty, the SPL transformation requires agent $i$ to additionally choose between $a_i^1$ and $a_i^2$ at the new information set $\bar{\boldsymbol{h}}_i$. SPL is a type of ``addition of a superfluous move'', a basic transformation in \cite{Thompson1952} \citep[also see,][]{Elmes1994,Wang2024}, applied to a subset of terminal histories of the original gradual mechanism.

\begin{figure}[ht]
  \centering
  \includegraphics[scale=0.85]{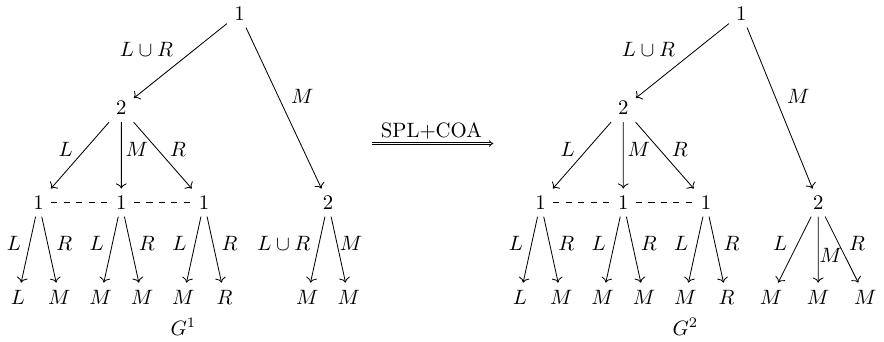}
  \caption{An example of SPL}
  \label{figSPL}
\end{figure}

Figure \ref{figSPL} exemplifies an SPL, accompanied by a COA transformation to be introduced shortly. In this figure, when $L$, $M$, and $R$ appear in available actions, they refer to the singleton sets of the three respective types. The SPL in this example is with respect to agent $2$'s action $L\cup R$ at the information set following agent $1$'s initial choice of $M$, splitting it into $L$ and $R$. We have the following proposition, the proof of which is in Appendix \ref{proofsec3}. 



\begin{proposition}
  Suppose a gradual mechanism $G^1$ can be transformed into another $G^2$ through an SPL. They implement the same SCF, and $G^1$ is incentive compatible if and only if $G^2$ is.
  \label{propsplic}
\end{proposition}

\subsubsection{Coalescing}

Let $\boldsymbol{h}_i$ and $\bar{\boldsymbol{h}}_i$ be two information sets in which $\bar{\boldsymbol{h}}_i$ immediately follows $\boldsymbol{h}_i$, i.e., there is no other information set of agent $i$ between them. If agent $i$ does not acquire additional information when arriving at $\bar{\boldsymbol{h}}_i$ after choosing $a_i$ at $\boldsymbol{h}_i$---any terminal history that passes through $\boldsymbol{h}_i$ where agent $i$ chooses $a_i\in A(\boldsymbol{h}_i)$ also passes through $\bar{\boldsymbol{h}}_i$---the COA transformation shifts the available actions at $\bar{\boldsymbol{h}}_i$ to $\boldsymbol{h}_i$, replacing the original action $a_i$. COA is also a basic transformation in \cite{Thompson1952}, where it is called ``coalescing of moves''.

\begin{figure}[ht]
  \centering
  \includegraphics[scale=0.85]{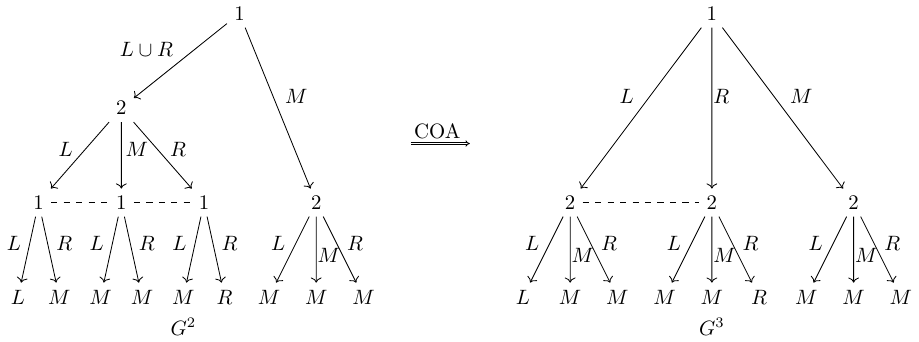}
  \caption{An example of COA}
  \label{figCOA}
\end{figure}

In $G^2$ on the left side of Figure \ref{figCOA}, voter $1$ has two information sets. Upon reaching the later information set, she learns nothing new about voter $2$'s preference after reporting $L \cup R$ at the former, initial information set. The COA advances the two finer reports $L$ and $R$ at the latter information set to replace $L \cup R$. Similar to SPL, we have the following proposition for COA, whose proof is also in Appendix \ref{proofsec3}.

\begin{proposition}
  Suppose a gradual mechanism $G^1$ can be transformed into another $G^2$ through a COA. They implement the same SCF, and $G^1$ is incentive compatible if and only if $G^2$ is.
  \label{propcoaic}
\end{proposition}

\subsubsection{Illuminating}

Let $\boldsymbol{h}_i$ be an information set in a gradual mechanism $G$. Let $\boldsymbol{h}_i^1$ and $\boldsymbol{h}_i^2$ be two non-empty disjoint subsets of $\boldsymbol{h}_i$ with their union being $\boldsymbol{h}_i$. After the ILL transformation, $\boldsymbol{h}_i^1$ and $\boldsymbol{h}_i^2$ are agent $i$'s information sets, replacing the original information set $\boldsymbol{h}_i$. Any information set $\bar{\boldsymbol{h}}_i$ in $G$ that is a successor of $\boldsymbol{h}_i$ needs to be partitioned accordingly to preserve perfect recall, i.e., into two information sets $\bar{\boldsymbol{h}}_i^1 = \{\bar{h}\in \bar{\boldsymbol{h}}_i: \exists h\in \boldsymbol{h}_i^1 \mbox{ such that } h\prec \bar{h}\}$ and $\bar{\boldsymbol{h}}_i^2 = \{\bar{h}\in \bar{\boldsymbol{h}}_i: \exists h\in \boldsymbol{h}_i^2 \mbox{ such that } h\prec \bar{h}\}$. Notice that one of these two sets may be empty, in which case $\bar{\boldsymbol{h}}_i$ is not meaningfully partitioned. The ILL transformation considered here is more fine-grained than that in \cite{Mackenzie2020} where all information sets are transformed into singletons at once.

\begin{figure}[ht]
  \centering
  \includegraphics[scale=0.85]{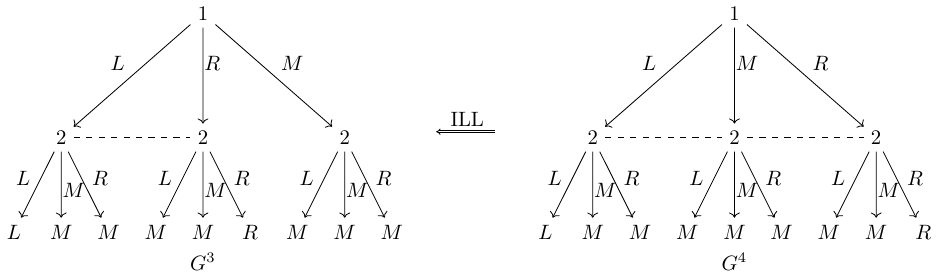}
  \caption{An example of ILL}
  \label{figILL}
\end{figure}

An ILL gives an agent additional scope to condition her action on her acquired information about other agents. For instance, $G^4$ in Figure \ref{figILL} is converted into $G^3$ through an ILL that partitions voter $2$'s information set, allowing her to distinguish whether voter $1$'s most preferred candidate is $M$. Consequently, voter $2$ could report differently after learning about whether voter $1$ has reported $M$ or not. For the gradual mechanism to be incentive compatible after an ILL partitioning $\boldsymbol{h}_i$ into $\boldsymbol{h}_i^1$ and $\boldsymbol{h}_i^2$, truth-telling for any other agent $j$ must remain dominant after taking into account this flexibility from agent $i$. Consider the following definition of an incentive-preserving ILL.


\begin{definition}
\label{def:IPILL}
  

  An ILL partitioning $\boldsymbol{h}_i$ into $\boldsymbol{h}_i^1$ and $\boldsymbol{h}_i^2$ is incentive-preserving if for any $\theta_i^1,\theta_i^2\in \Theta_i(\boldsymbol{h}_i)$, any $\theta_j^1,\theta_j^2\in \Theta_j$, and any $\theta_{-i,j}^1, \theta^2_{-i,j}\in \Theta_{-i,j}$ such that (i) $(\theta_j^1, \theta_{-i,j}^1)\in \Theta_{-i}(\boldsymbol{h}_i^1)$, (ii) $(\theta_j^2, \theta_{-i,j}^2)\in \Theta_{-i}(\boldsymbol{h}_i^2)$, and (iii) there exists $s_{-i,j} \in S_{-i,j}=\prod_{k \in N \setminus \{i,j\}} S_k$ consistent with both $(\theta_i^1, \theta_j^1, \theta_{-i,j}^1)$ and $(\theta_i^2, \theta_j^2, \theta_{-i,j}^2)$, it is the case that $f(\theta_i^1, \theta_j^1, \theta_{-i,j}^1) R(\theta_j^1) f(\theta_i^2, \theta_j^2, \theta_{-i,j}^2)$.


  
  

\end{definition}

In this definition, $\theta_i^1$ and  $\theta_i^2$ could belong to different available actions at $\boldsymbol{h}_i$, capturing agent $i$'s flexibility to condition her actions on the different information she acquires. The three premises stipulate that when agents other than $i$ and $j$ are choosing according to $s_{-i,j}$, agent $j$'s action---consistent with $\theta_j^1$ or $\theta_j^2$---determines which information set---$\boldsymbol{h}_i^1$ or $\boldsymbol{h}_i^2$---obtains. To put it differently: under the speculation $s_{-i,j}$ about other agents' strategies, agent $i$ learns that agent $j$'s type might be $\theta_j^1$ at $\boldsymbol{h}_i^1$ and might be $\theta_j^2$ at $\boldsymbol{h}_i^2$. Incentive-preserving ILL requires that, in the face of agent $i$'s additional flexibility, truthfully revealing her type between $\theta_j^1$ and $\theta_j^2$ remains agent $j$'s dominant strategy.\footnote{The three premises hold after partially switching the superscripts $1$ and $2$, therefore the incentive-preserving condition on ILL also requires $f(\theta_i^2, \theta_j^1, \theta_{-i,j}^1) R(\theta_j^1) f(\theta_i^1, \theta_j^2, \theta_{-i,j}^2)$, $f(\theta_i^1, \theta_j^2, \theta_{-i,j}^2) R(\theta_j^2) f(\theta_i^2, \theta_j^1, \theta_{-i,j}^1)$, and $f(\theta_i^2, \theta_j^2, \theta_{-i,j}^2) R(\theta_j^2) f(\theta_i^1, \theta_j^1, \theta_{-i,j}^1)$.}

Definition \ref{def:IPILL} resembles Li's (2017) obvious dominance from the perspective that both require the dominant strategy to deliver (weakly) better outcomes even when other agents are choosing differently against the dominant strategy and an alternative strategy.\footnote{Indeed, if truth-telling is obviously dominant in a gradual mechanism, then all of its potential ILLs are incentive-preserving. \cite{Mackenzie2020} relies on a similar observation to provide a revelation principle for obviously dominant strategy implementation.} Note that the ILL in Figure \ref{figILL} is incentive-preserving. We prove the following proposition in Appendix \ref{proofsec3}.


\begin{proposition}
  \label{IP}
  Suppose a gradual mechanism $G^1$ can be transformed into another $G^2$ through an ILL. They implement the same SCF, and $G^2$ is incentive compatible if and only if $G^1$ is incentive compatible and the ILL is incentive-preserving.
  \label{propillic}
\end{proposition}

\subsubsection{Characterization}

Under the assumption (made for convenience) that the initial history is a (possibly degenerate) information set for each agent, the three basic transformations can be applied sequentially to transform any gradual mechanism into the direct mechanism of the same SCF. For instance, with one SPL, two COAs, and an inverse ILL, the gradual mechanism $G^1$ in Figure \ref{figSPL} is converted into $G^4$ in Figure \ref{figILL}. A subsequent COA, which advances voter 2's decision to the initial history, then transforms $G^4$ into the direct mechanism.

\begin{proposition}
\label{Reduction}
  Any gradual mechanism $G$ implementing an SCF $f$ can be transformed into the direct mechanism of $f$ through a chain of SPLs, COAs, and inverse ILLs.
\end{proposition}

Combining propositions \ref{propsplic}-\ref{Reduction}, we derive the following theorem and omit the proof.

\begin{theorem}
  \label{theotrans}
  A gradual mechanism $G$ implementing a strategy-proof SCF $f$ is incentive compatible if and only if, in some chain of SPLs, COAs, and inverse ILLs that transforms $G$ into the direct mechanism implementing $f$, all ILLs are incentive-preserving.
  \label{theo1}
\end{theorem}

This characterization is of particular interest. ILL is the fundamental transformation that sequentializes agents' decision-making processes by providing them incremental information, and is thus the route by which gradual mechanisms become strategically simpler, less intrusive on privacy, or more credible. This collection of basic transformations provides a toolkit for the designer to start with the direct mechanism and search for a desirable incentive compatible gradual mechanism that implements a given SCF. Moreover, as will be illustrated in our application to ascending-price auctions, the incentive-preservation condition on ILLs is useful to show when a maximal level of information transmission among the agents has been achieved.



Indeed, Theorem \ref{theotrans} suggests how demanding it is for a particular gradual mechanism to be incentive compatible, as a whole sequence of ILLs must be incentive-preserving. This raises a natural question: how much scope remains for designing incentive compatible dynamic mechanisms? Say that a gradual mechanism $G$ is \textit{non-static} if there exists some agent $i$ who learns some information about other agents' types at some information set $\boldsymbol{h}_i$, i.e., $\Theta_{-i}(\boldsymbol{h}_i) \subsetneq \Theta_{-i}$. The following proposition characterizes an extreme scenario when a strategy-proof SCF does not admit any incentive compatible non-static gradual mechanism.

\begin{proposition}
  \label{propscf}
  There is no incentive compatible non-static gradual mechanism implementing a strategy-proof SCF $f$ if and only if for any $i \in N$, any distinct $\theta^1_i,\theta^2_i \in \Theta_i$, and any non-trivial partition $\{R^1_{-i},R^2_{-i}\}$ of $\Theta_{-i}$, there exist $j \in N$ with $j \neq i$, $(\theta^1_j,\theta_{-i,j}) \in R^1_{-i}$ and $(\theta^2_j,\theta_{-i,j}) \in R^2_{-i}$ such that for some $k,l \in \{1,2\}$, it is the case that $f(\theta^k_j,\theta^l_i,\theta_{-i,j})P(\theta^{3-k}_j)f(\theta^{3-k}_j,\theta^{3-l}_i,\theta_{-i,j})$.
\end{proposition}

The characterization pictures a minimal opportunity for agent $i$ to break other agents' incentives. It is when agent $i$ can only condition her reports between two types $\theta^1_i$ and $\theta^2_i$ on a minimal amount of information about other agents captured by a binary partition $\{R^1_{-i},R^2_{-i}\}$ of $\Theta_{-i}$. If any such minimal opportunity breaks some agent $j$'s incentive, no non-static gradual mechanism can implement $f$ in dominant strategies. Conversely, if all agents' truthful incentives survive such a minimal opportunity, then a non-static gradual mechanism can be constructed from the direct mechanism implementing $f$. This is done through an inverse COA that allows agent $i$ to delay distinguishing between $\theta_i^1$ and $\theta_i^2$ until other agents have reported their exact types, followed by an incentive-preserving ILL that enables $i$ to learn whether the others' types lie in $R_{-i}^1$ or $R_{-i}^2$. See Appendix \ref{proofsec3} for the proof.

\subsection{Reaction-Proofness}

The preceding analysis inspires the following reaction-proofness condition. This condition is local: it examines only how an agent can react to another agent immediately after acquiring information through a pair of (distinct) information sets following the same action at a common immediate predecessor.

\begin{definition}
  \label{def:RP}
  A gradual mechanism $G$ implementing an SCF $f$ is reaction-proof if for any two agents $i, j\in N$, any pair of agent $i$'s information sets $\boldsymbol{h}_i^1$ and $\boldsymbol{h}_i^2$ that are immediate successors of a common $\boldsymbol{h}_i$ with $\Theta_i(\boldsymbol{h}_i^1) = \Theta_i(\boldsymbol{h}_i^2)$, any pair of histories $h^1\in \boldsymbol{h}_i^1$ and $h^2\in \boldsymbol{h}_i^2$ consistent with a common strategy profile $s_{-i,j} \in S_{-i,j}$, it is the case that $f(\theta^1) R(\theta_j^1) f(\theta^2)$ for any pair of type profiles $\theta^1\in \Theta(h^1)$ and $\theta^2\in \Theta(h^2)$ that are consistent with the same $s_{-i,j}$.


\end{definition}

Definitions \ref{def:IPILL} and \ref{def:RP} are similar, as are their interpretations. Under the speculation $s_{-i,j}$ about other agents' strategies, agent $i$ learns that $h^1 \in \boldsymbol{h}^1_i$ (or $h^2 \in \boldsymbol{h}^2_i$) results from $j$ reporting type $\theta^1_j$ (or $\theta^2_j$).\footnote{Since $s_{-i,j}$ is fixed and agent $i$ encounters the same sequence of information sets before $\boldsymbol{h}^1_i$ and $\boldsymbol{h}^2_i$, agent $j$'s different actions at a common predecessor of $h^1$ and $h^2$ lead to the first divergence of the paths from the initial history to $h^1$ and $h^2$.} Reaction-proofness requires that, regardless of agent $i$'s reaction at $h^1$ and $h^2$, agent $j$ has an incentive to truthfully reveal her type between $\theta^1_j$ and $\theta^2_j$. Note that reaction-proofness can be further relaxed by requiring $f(\theta^1) R(\theta_j^1) f(\theta^2)$ to hold only for agent $i$ who is among the first to acquire information about agent $j$'s choice under the speculation $s_{-i,j}$.\footnote{Formally, this can be done by further requiring that there are no distinct information sets $\boldsymbol{h}_k^1$ and $\boldsymbol{h}_k^2$ of some agent $k$ other than $i$ and $j$ such that there exist $\underline{h}^1\in \boldsymbol{h}_k^1$ and $\underline{h}^2\in \boldsymbol{h}_k^2$ with $\underline{h}^1\prec h^1$ and $\underline{h}^2\prec h^2$. This also applies to the following strong reaction-proofness.} We prove the following theorem in Appendix \ref{proofsec3}.\footnote{To prove Theorem \ref{theorp} using Theorem \ref{theotrans}, we can show that a gradual mechanism $G$ is reaction-proof if and only if there is a chain of SPLs, COAs, and inverse ILLs transforming $G$ into the direct mechanism in which all the ILLs are incentive-preserving. This can be done because, like incentive compatibility, only incentive-breaking ILLs change the reaction-proofness of two gradual mechanisms linked by them. We adopt a more direct approach to prove Theorem \ref{theorp} based on the intuition to be discussed below.}


\begin{theorem}
  \label{theorp}
  A gradual mechanism implementing a strategy-proof SCF is incentive compatible if and only if it is reaction-proof.
  
\end{theorem}

\begin{figure}[ht]
  \centering
  \includegraphics[scale=0.99]{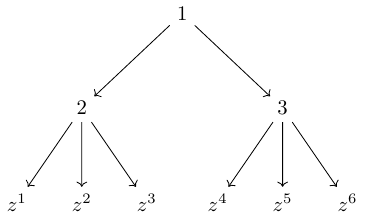}
\caption{A gradual mechanism vacuously satisfying reaction-proofness}
\label{fig:RP}
\end{figure}

Theorem \ref{theorp} indicates that, to show incentive compatibility of a gradual mechanism, we can safely restrict our attention to its structures where one agent can react to another agent, thus omitting all of its other features. For instance, the gradual mechanism in Figure \ref{fig:RP}, where reaction-proofness holds vacuously because no agent can react to any other, must be incentive compatible as long as its implemented SCF is strategy-proof.

As mentioned in the previous subsection, the condition of incentive-preserving ILL shares the spirit of Li's (2017) obvious dominance. We now illustrate this point in the following application of reaction-proofness to another simple voting setting. This example also points out how we can strengthen reaction-proofness by applying the idea of relaxing contingent reasoning, which motivates Li's (2017) concept.

\begin{figure}[ht]
  \centering
  \includegraphics{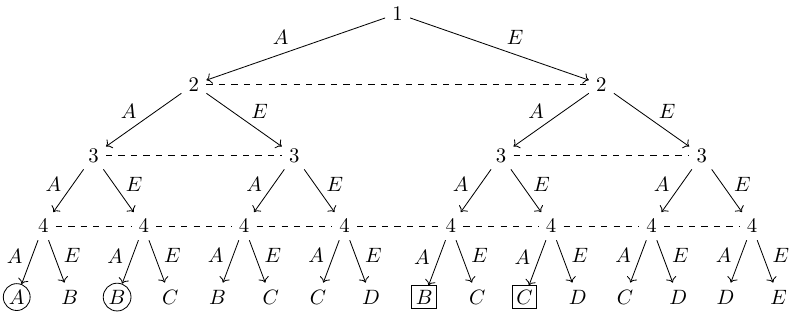}
\caption{An example of reaction-proofness}
\label{fig:SRP}
\end{figure}

Consider a situation where four voters in $N = \{1, 2, 3, 4\}$ choose one candidate from five candidates in $X = \{A, B, C, D, E\}$. Each voter is either of type $A$ or of type $E$. A type $A$ voter strictly prefers candidate $A$ over $B$ over $C$ over $D$ and over $E$, while a type $E$ voter has the reversed preference. Consider a voting rule $f$ that selects $A$ (or $E$, respectively) if all voters are of type $A$ (or type $E$), and selects $B$ (or $D$) if three voters are of type $A$ (or type $E$), and select $C$ if two voters are of type $A$, which is strategy-proof. Finally, consider the gradual mechanism $G$ in Figure \ref{fig:SRP} implementing the above SCF $f$, where voters reveal their types sequentially, and voter $3$ observes voter $1$'s choice.

It is obvious that $G$ only allows voter $3$ to react to voter $1$. Therefore, to show $G$ is incentive compatible via its reaction-proofness, it suffices to examine the condition with respect to how voter $1$'s incentive could be compromised by voter $3$'s additional information. Let $\boldsymbol{h}^1_3$ and $\boldsymbol{h}^2_3$ be voter $3$'s information sets following voter $1$ reporting $A$ and $E$, respectively. Now suppose voter $1$ is considering the contingency that both voter $2$ and voter $4$ report $A$. This leads voter $1$ to think of $h^1 = \langle A, A\rangle \in \boldsymbol{h}^1_3$ and $h^2 = \langle E, A\rangle \in \boldsymbol{h}^2_3$ which are consistent with the contingency. The terminal histories following $h^1$ and $h^2$ that are also consistent with the contingency are surrounded with circles and rectangles in the figure, respectively. It is immediate to see that for voter $1$ of type $A$, outcomes in circles are better than or equal to outcomes in rectangles, and for voter $1$ of type $E$, outcomes in rectangles are better than or equal to outcomes in circles. To confirm that reaction-proofness holds, voter $1$ thinks through all the possible contingencies of the choices of voter $2$ and voter $4$.

As we can see from the above analysis, the application of reaction-proofness requires voter $1$ to engage in complex contingent reasoning. For each contingency of voter $2$ and voter $4$'s strategic choice, voter $1$ not only conceives a pair of histories where voter $3$ could react to her, but also thinks of the outcomes that are consistent with the contingency. One way to reduce the burden of contingent reasoning is to drop the requirement of identifying consistent outcomes. In this spirit, we provide the following strong reaction-proofness.

\begin{definition}
  A gradual mechanism $G$ implementing an SCF $f$ is strongly reaction-proof if for any two agents $i, j\in N$, any pair of agent $i$'s information sets $\boldsymbol{h}_i^1$ and $\boldsymbol{h}_i^2$ that are immediate successors of a common $\boldsymbol{h}_i$ with $\Theta_i(\boldsymbol{h}_i^1) = \Theta_i(\boldsymbol{h}_i^2)$, for any pair of histories $h^1\in \boldsymbol{h}_i^1$ and $h^2 \in \boldsymbol{h}_i^2$ consistent with a common strategy profile $s_{-i,j} \in S_{-i,j}$, there exists $h^k\in \{h^1, h^2\}$ such that $f(\theta)R(\theta^*_j)f(\theta')$ for any $\theta, \theta'\in \Theta(h^k)$ and any $\theta^*_j\in \Theta_j$.
  \label{sufficientrp}
\end{definition}

Strong reaction-proofness requires that whenever agent $i$ can react to agent $j$ at two histories, agent $j$ is indifferent among all outcomes following at least one of them. Hence, once agent $j$ envisions such a pair of histories, she needs only compare the outcomes she finds indifferent following one history with the best (or worst) outcome following the other to see that truth-telling is dominant. For instance, the gradual mechanism $G^3$ in Figure \ref{figILL} is strongly reaction-proof because, once voter $1$ reports $M$ at the initial information set, the winning candidate $M$ is already determined. 

In Appendix \ref{proofsec3}, we prove the sufficiency as stated in the following theorem by showing that strong reaction-proofness, together with strategy-proofness of the implemented SCF, implies reaction-proofness.

\begin{theorem}
\label{theo:IRP}
  If a gradual mechanism implementing a strategy-proof SCF is strongly reaction-proof, then it is reaction-proof.
\end{theorem}

As discussed in the introduction, our strong reaction-proofness adapts the reaction-proofness condition of menu mechanisms in \cite{mackenzie2022}. Their reaction-proofness is defined via identical assignments, without reference to SCFs or agents' preferences, and is part of a sufficient condition for truth-telling to be everywhere dominant. In contrast, our notion extends this condition to gradual mechanisms, applying in more general environments without assumptions on preferences except for completeness and transitivity.

\section{Applications}
\label{secapp}
\subsection{An Application to Ascending-Price Auctions}
\label{secauction}
We consider a dynamic mechanism design problem where an indivisible item is allocated to one of several bidders according to the second-price auction rule. Moreover, to enhance strategic simplicity and privacy protection, the auctioneer is considering the format of ascending-price auctions, some of which involve information transmission within the same price level in an environment with discrete private values.


If the bidding information is openly disseminated during the auction, suspicions of spiteful bidding, i.e., a bidder staying in the auction merely to raise another bidder's payment, might harm the incentive compatibility of the resulting ascending-price auction. However, concealing such information might be costly or impractical and harm the credibility of the auction. Thus, the auctioneer might want to identify the boundary of information transmission among bidders without compromising incentive compatibility, motivating us to investigate \textit{transparent} ascending-price auctions in the following sense.\footnote{\cite{nagel2024if} define a transparency order on mechanisms with isomorphic histories based on the coarseness of agents' information sets. An incentive compatible gradual mechanism is transparent in our sense if any gradual mechanism that is more transparent in their sense is not incentive compatible.}

\begin{definition}
  An incentive compatible gradual mechanism is transparent if there is only one active agent at each non-terminal history and none of its potential ILLs is incentive-preserving.
  \label{defi:transparency}
\end{definition}

Specifically, consider an auction setting in which each bidder $i\in N = \{1, \ldots, n\}$ has a private value $v_i\in V = \{1, \ldots, m\}$ for the auctioned item. A social outcome consists of a stochastic allocation rule of the item to bidders, along with the prices they need to pay if they win the auction. A bidder's preference over different social outcomes is determined by her expected payoff, i.e., her winning probability times the difference between her valuation of the item and her winning price. Given a profile $v \in  V^n$ of private values, the second-price auction rule delivers a social outcome that allocates, with equal probability, the item to one of the agents with the highest value for it who pays the price equal to the second-highest value. It is well-known that the second-price auction rule is strategy-proof.

A wide range of ascending-price auctions can be applied to implement the second-price auction rule. The core procedure in common is the following.\footnote{We restrict attention to deterministic game forms, allowing randomization only at terminal histories.}

\begin{definition}
In an ascending-price auction, starting with $1$, at each price level $p < m$, the remaining bidders choose to stay in or leave the auction, and:
\begin{enumerate}
\item if two or more  bidders stay at $p$, increase the price level by $1$;
\item if only one bidder stays at $p$, she wins and pays $p$;
\item if all leave at $p$, a random bidder among them wins and pays $p$;
\item if the price reaches $m$, a random remaining bidder wins and pays $m$.
\end{enumerate}
\end{definition}

Let bidder $i$'s decision to stay in the auction at price level $p$ correspond to $\{v_i\in V: v_i > p\}$ and leaving the auction at $p$ correspond to $\{v_i\in V: v_i = p\}$. It is straightforward to see that these ascending-price auctions are a class of gradual mechanisms, differing from each other in terms of the information each bidder acquires when making stay/leave decisions at various price levels. At one extreme, bidders only know that the auction has not yet terminated. At the other end, there are perfect information ascending-price auctions.

It is intuitive that information about every bidder's decisions at previous price levels can be provided to the remaining bidders without harming their incentives \citep[e.g.,][]{krishna2009auction}. For the transmissibility of information generated within the same price level, consider the following two simple examples where there are only two price levels.

\begin{example}
\label{example1}
Suppose $n = 2$ and $m = 2$. Let bidder $2$ be informed about bidder $1$'s choice at price level $1$. Then, for bidder $1$ with private value $2$, truth-telling is not a dominant strategy. To see this, suppose bidder $2$'s strategy is to leave the auction after learning bidder $1$ has left and to stay in the auction after learning bidder $1$ has stayed. Then, leaving the auction at price level $1$ delivers bidder $1$ an expected payoff of $0.5$, greater than the expected payoff of $0$ by staying in the auction.
\end{example}

Example \ref{example1} shows that illuminating some information about a bidder's decision at the current price level may destroy her truthful incentive. The following Example \ref{example2} demonstrates that it is not always the case, i.e., some information can indeed be transmitted without hurting the incentive for truth-telling.

\begin{example}
\label{example2}
Suppose $n = 3$ and $m = 2$. Let bidders $1$ and $2$ make decisions simultaneously. Let bidder $3$ be informed about whether or not bidders $1$ and $2$ have both stayed in the auction. This ascending-price auction is essentially delivered by a single ILL from the direct mechanism. We can show that it is incentive compatible since this ILL is incentive-preserving. To see this, first apply the condition with respect to bidders $1$ and $3$. Bidder $3$ might learn bidder $1$'s decision under the speculation that bidder $2$ has stayed in the auction. Note that staying in the auction corresponds to bidding $2$ and leaving the auction corresponds to bidding $1$. Let $EV(b_1, b_2, b_3 | v_1)$ where $b_1, b_2, b_3, v_1\in \{1, 2\}$ be bidder $1$'s expected payoff given bids $(b_1, b_2, b_3)$ when her true private value is $v_1$. The incentive-preservation condition requires us to examine the following four inequalities: $EV(2, 2, 1 | 2)\geq EV(1, 2, 2 | 2)$, $EV(2, 2, 2 | 2)\geq EV(1, 2, 1 | 2)$, $EV(1, 2, 2 | 1)\geq EV(2, 2, 1 | 1)$, and $EV(1, 2, 1 | 1)\geq EV(2, 2, 2 | 1)$. Since all four inequalities hold, truth-telling is dominant for bidder $1$. The dominance of truth-telling for bidder $2$ can be similarly demonstrated.
\end{example}

The application of the incentive-preservation condition in Example \ref{example2} suggests the following hypothesis: an ascending-price auction is incentive compatible if and only if, at each price level $p$, no bidder $i$, holding the unrefuted speculation that all remaining bidders leave the auction at $p$ except for a specific bidder $j$, could know whether $j$ stays or leaves. To capture this idea about the transmissibility of information within the same price level, we define the following \textit{open-by-two ascending-price auction}, in which all actions at the current price level become public once two bidders choose to stay.

\begin{definition}
An open-by-two ascending-price auction is such that:
\begin{enumerate}
  \item a bidder with a smaller index chooses earlier at each price level;
  \item each bidder observes all the actions at previous price levels;
  \item each bidder observes all the actions at the current price level $p$ if at least two bidders have chosen to stay at $p$; otherwise, she is informed that at most one bidder has chosen to stay at $p$.
\end{enumerate}
\end{definition}

\begin{figure}[ht]
  \centering
  \includegraphics{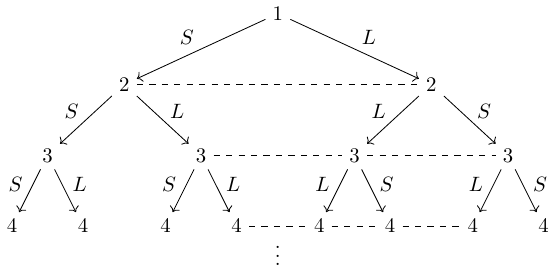}
\caption{The first few steps of the open-by-two ascending-price auction}
\label{fig:auction}
\end{figure}

Figure \ref{fig:auction} presents the first few steps of the open-by-two ascending-price auction, in which $S$ and $L$ refer to stay in and leave the auction, respectively. In addition to confirming the intuition that all information from previous price levels can be made public, the following proposition also validates that within the same price level, only the aforementioned information transmission can be done without harming incentive compatibility.

\begin{proposition}
  \label{propapaic}
  The open-by-two ascending-price auction is incentive compatible. Moreover, it is transparent.
\end{proposition}

The open-by-two ascending-price auction is strongly reaction-proof. Suppose bidder $i$ can react to bidder $j$ at price $p$ at a pair of histories $h^1$ and $h^2$. As the first divergence of $h^1$ and $h^2$ results from bidder $j$'s different actions at a common predecessor, bidder $j$ must have left the auction before either $h^1$ or $h^2$, say $h^1$. It is clear that bidder $j$ has lost the auction at $h^1$ if the decision to leave is made at price $p - 1$. If bidder $j$ left the auction at price $p$ along the path to $h^1$ (while she stayed along the path to $h^2$), there must be another bidder $k$ with $k<i$ who stayed at price $p$ along the paths to both $h^1$ and $h^2$ without learning about bidder $j$'s decision. In this case, bidder $j$ has also lost at $h^1$. See Appendix \ref{proofsec4} for the proof of Proposition \ref{propapaic}.

\subsection{An Application to Hierarchical Exchange Rules}
\label{secrda}
Consider a finite set $O$ of objects to be allocated to a finite set $N$ of agents. An allocation is a function $x: N \to O \cup \{\varnothing\}$ such that $|x^{-1}(o)| \leq 1$ for any $o\in O$. $x(i) \in O$ means agent $i$ is assigned object $x(i)$ under $x$, while $x(i)=\varnothing$ means agent $i$ is not assigned any object. Let $X$ denote the set of allocations. For $N' \subseteq N$ and $O' \subseteq O$ with $\lvert N' \rvert=\lvert O' \rvert \neq 0$, let $X(N',O')$ denote the set of bijections from $N'$ to $O'$, whose elements are called \textit{partial allocations}. In particular, define $X(\varnothing,\varnothing)=\{\varnothing\}$.

Let $\mathcal{P}$ be the set of complete, transitive, and strict preference relations over $O \cup \{\varnothing\}$ such that each object in $O$ is preferred to $\varnothing$. Assume that for each agent $i$, there is a bijection $P: \Theta_i \to \mathcal{P}$. Let $R(\theta_i)$ denote the weak preference induced by $P(\theta_i)$, i.e., $\forall o,o' \in O$, $o R(\theta_i) o'$ if and only if $o=o'$ or $o P(\theta_i) o'$. We suppose no consumption externalities, i.e., an allocation $x \in X$ is preferred to $x'$ by type $\theta_i$ if $x(i) R(\theta_i) x'(i)$, also written as $x R(\theta_i) x'$.

It is well-known that in such an environment, hierarchical exchange rules (HERs) proposed by \cite{papai2000strategyproof} are a class of strategy-proof SCFs. For any type profile $\theta$, an HER $f^{\Gamma}$ determines an allocation $f^{\Gamma}(\theta)$ through the TTC algorithm working in stages based on a flexible inheritance structure $\Gamma$. In each stage $t$ with partial allocation $x^t$ realized in previous stages, according to the inheritance structure $\Gamma$, every remaining (i.e., unassigned) agent $i$ is endowed with a \textit{non-decreasing} menu $\mathcal{E}_i^\Gamma(x^t) \subseteq O$ of remaining objects and each object is owned by exactly one remaining agent.\footnote{Formally, for any $N' \subseteq N$ and $O' \subseteq O$ with $|N'|=|O'|$, any $x' \in X(N',O')$: (i) $\{\mathcal{E}_i^\Gamma(x') \neq \varnothing: i \in N \setminus N'\}$ is a partition of $O \setminus O'$, and (ii) for any extension $x'' \in X(N'',O'')$ of $x'$ and any $i \in N \setminus N''$, 
$\mathcal{E}_i^\Gamma(x') \subseteq \mathcal{E}_i^\Gamma(x'')$. The menu $\mathcal{E}_i^\Gamma(\cdot)$ can be defined by a profile of the so-called inheritance trees $\Gamma=(\Gamma_o)_{o \in O}$. See \cite{papai2000strategyproof} and \cite{mandal2022obviously} for the formal definition of inheritance trees and how they determine ownership of objects in the algorithm of HER.} Each agent points to her most preferred object, and each object points to its owner, thereby inducing a directed graph. This graph contains at least one cycle, and the agents in each cycle receive the objects they point to. The next stage begins after removing the assigned agents and objects from the market, and the algorithm terminates once all agents or all objects have been assigned.

We study the incentive compatibility of a class of gradual mechanisms, called renunciation-designation-assertion (RDA) mechanisms. These mechanisms were first proposed by \citet{CW2024} for the TTC algorithm to reduce agents' burden of contingent reasoning, extending the dynamic mechanisms in \cite{Troyan2019} and \cite{mandal2022obviously} to environments where obviously dominant strategy implementations are unavailable. We generalize RDA mechanisms to implement the HER. Unlike some dynamic mechanisms that always ask agents about their favorite objects from given menus \citep[e.g.,][]{mackenzie2022,bo2024pick,nagel2025measure}, paralleling the TTC algorithm, RDA mechanisms mainly inquire agents about the menu containing their favorite object among a given collection of menus. This greater departure from the TTC algorithm makes the proof of the incentive compatibility of RDA mechanisms more demanding.

An RDA mechanism works in stages, producing partial allocations through three sub-stages: renunciation, designation, and assertion. For each remaining agent, it tracks a designated trading partner, an endowed menu of objects that she owns according to the inheritance structure of the HER being implemented and the partial allocation already realized, and a tentative menu of objects from which she may choose upon entering the assertion sub-stage. Initially, no trading partner is designated for any agent, and each agent's tentative menu is the set of all objects.

Each stage begins with a renunciation sub-stage, in which some standby owners—specifically, those with non-empty endowed menus and no trading partners—are called sequentially to decide whether to renounce their own objects. If such an owner decides to renounce, her tentative menu is updated to the objects outside her endowed menu; otherwise, her new tentative menu is the intersection of her current tentative menu and her endowed menu. The renunciation sub-stage ends either when some owner $i$ decides not to give up her objects or when all such owners make renunciations. In the former case, $i$ forms a singleton trading cycle, triggering the assertion sub-stage. In the latter case, the designation sub-stage begins.

In the designation sub-stage, these standby owners are called sequentially. Each active standby owner chooses one endowed menu, owned by another owner, that has a non-empty intersection with her tentative menu; by choosing that menu, she designates its owner as her trading partner. When owner $i$ designates owner $j$ as a trading partner in this sub-stage, owner $i$'s tentative menu is updated to its intersection with owner $j$'s endowed menu. Designations persist unless the designated partner exits. The designation sub-stage terminates, and the assertion sub-stage begins, once a multi-agent trading cycle emerges.

In the assertion sub-stage, each owner in the newly formed trading cycle simultaneously chooses an object from her tentative menu. These agents and objects are then removed from the market. When a removed object belonged to the tentative menu of a remaining agent, that agent's tentative menu is reset to the set of all remaining objects; otherwise, tentative menus are left unchanged. The mechanism stops once all agents or all objects have been assigned.

Formally, given an HER $f^{\Gamma}$ identified by inheritance structure $\Gamma$, the RDA mechanism for $f^{\Gamma}$ works as follows. 

\medskip

\noindent
\textbf{Definition of RDA mechanism.} The mechanism proceeds in stages with renunciation, designation, and assertion sub-stages, starting with stage $1$. Agents choose tentative menus repeatedly in renunciation and designation sub-stages until they enter an assertion sub-stage and leave the mechanism with objects chosen from their tentative menus. For a positive integer $t$, let $h^t$ be an initial history of stage $t$; denote the sets of matched agents and objects up to $h^t$ as $N(h^t)$ and $O(h^t)$, respectively; and let $x^t \in X(N(h^t),O(h^t))$ be the partial allocation made before $h^t$. Next, we describe how the mechanism proceeds in stage $t$ following $h^t$, first introducing the notions of endowed menus, tentative menus, and standby owners.


\medskip

\textit{Endowed Menus.}
Each remaining agent $i \in N \setminus N(h^t)$ is endowed with a (possibly empty) menu $E_i(h^t) = \mathcal{E}^{\Gamma}_i(x^t)$, which is invariant within stage $t$.

\textit{Tentative Menus.}
For each remaining agent $i\in N \setminus N(h^t)$, at any history $h$ of stage $t$ following $h^t$, denote by $a^{-1}_i(h)$ her most recently chosen menu (which could have been chosen at a previous stage; let $a^{-1}_i(h)=O \setminus O(h^t)$ if $i$ has never acted before $h$). Agent $i$'s tentative menu $M_i(h)$ at $h$ is given by 

$$M_i(h)=
\begin{cases}
  a^{-1}_i(h), & \text{if } a^{-1}_i(h)\subseteq O \setminus O(h^t),\\
  O \setminus O(h^t), & \text{otherwise.}
\end{cases}.$$ Notice that, by this definition, an agent's tentative menu is updated upon arriving at $h^t$ and by the agent's own actions during stage $t$.

\textit{Standby owners.} At any history, we say an agent has no trading partner if she has never designated a partner (see the definition below), or her most recently designated partner has left the market. Say an agent $i\in N\setminus N(h^t)$ is a standby owner in stage $t$ if (i) $E_i(h^t)\neq\varnothing$ and (ii) $i$ has no trading partner at $h^t$.

\medskip
Stage $t$ begins with a renunciation sub-stage.

\medskip
\noindent
\textit{1. Renunciation sub-stage.} Standby owners take actions in increasing index order. At a history $h$ of stage $t$ following $h^t$ where standby owner $i$ is active, she chooses between two menus $M_i(h)\setminus E_i(h^t)$ and $M_i(h)\cap E_i(h^t)$ partitioning her tentative menu with her endowed menu; choosing the former means renouncing her endowed menu.\footnote{In either case, if one of the two menus is empty, the decision node is skipped for $i$, and $i$ is treated as choosing the other non-empty menu.} The renunciation sub-stage terminates if (i) all standby owners make renunciations, or (ii) a standby owner $j$ does not renounce. In the former case, the designation sub-stage begins. In the latter case, a singleton trading cycle $C^{t}=\{j\}$ forms, and the assertion sub-stage begins.

\medskip
\noindent
\textit{2. Designation sub-stage.} Standby owners take actions in increasing index order. At a history $h$ of stage $t$ following $h^t$ where standby owner $i$ is active, she chooses among the non-empty menus in $\{M_i(h)\cap E_j(h^t):j\in N\setminus N(h^t),\ j\neq i\}$ which partition her current tentative menu with endowed menus of other agents; choosing $M_i(h)\cap E_j(h^t)$ means designating $j$ as her trading partner.\footnote{If only one menu $M_i(h)\cap E_j(h^t)$ is non-empty, the decision node is skipped for $i$, and $i$ is treated as choosing the only menu and designating $j$ as trading partner.} The designation sub-stage ends once a multiple-agent trading cycle $C^{t}=\{j_1,...,j_g\}$ forms, in which agent $j_s$ designates agent $j_{s+1}$ as her trading partner for each $s\in \{1, \ldots g - 1\}$ and agent $j_g$ designates agent $j_1$, and the assertion sub-stage begins.

\medskip
\noindent
\textit{3. Assertion sub-stage.} Each owner in the trading cycle $C^{t}$ simultaneously picks an object from her tentative menu and receives it.

\medskip
Stage $t+1$ begins after removing matched agents and objects from the market. If agent $i$ designates agent $j$ as her trading partner and agent $j$ is assigned an object in the assertion sub-stage of stage $t$, then agent $i$ is considered without a trading partner upon entering stage $t+1$. The mechanism terminates when all agents obtain an object or when all objects are assigned to an agent.

\medskip

\textit{Information structure.}
Agents observe only the available actions. Formally, for any $i\in N$, two decision nodes $h^1,h^2\in H_i$ are in the same information set if and only if there exists a bijection $m:\{h\in H_i:h\preceq h^1\}\to \{h\in H_i:h\preceq h^2\}$ such that (i) for any $h \in H_i$ with $h \preceq h^1$, $A_i(h)=A_i(m(h))$; (ii) for any $h,\bar{h} \in H_i$ with $h \prec \bar{h} \preceq h^1$, $m(h) \prec m(\bar{h})$; and (iii) for any $h \in H_i$ with $h \prec h^1$, the action chosen by $i$ at $h$ on the path to $h^1$ coincides with the action chosen by $i$ at $m(h)$ on the path to $h^2$.  \hfill $\square$

\medskip

At any decision node $h$ of any agent $i$ belonging to a renunciation or a designation sub-stage, the choice of a new tentative menu $M$ is understood to mean that $M$ contains agent $i$'s most preferred object among the remaining objects. Two design features of RDA mechanisms ensure that tentative menus work as intended. First, the collection of available new tentative menus always forms a partition of the current tentative menu. Second, when any object in agent $i$'s current tentative menu is assigned to another agent in an assertion sub-stage, agent $i$'s tentative menu is reset to the set of all remaining objects upon entering a new stage. These two rules guarantee that agent $i$ can never contradict her own previous reports and always has a truth-telling option. Thus, RDA mechanisms are gradual mechanisms, and their incentive compatibility can be analyzed using our characterizations. The result is affirmative, as stated in the following proposition.

\begin{proposition}
  \label{proprda}
  The RDA mechanism for any HER $f^{\Gamma}$ is incentive compatible and implements $f^{\Gamma}$.
  \end{proposition}

The RDA mechanism is designed so that an agent can affect other agents' subsequent decision-making only by leaving the market, a feature making strong reaction-proofness hold. Suppose agent $i$ can react to agent $j$ at a pair of histories $h^1$ and $h^2$ that are consistent with a common strategy profile $s_{-i,j}$. It must be the case that agent $j$'s different actions at a common predecessor lead the paths to $h^1$ and $h^2$ to diverge, while agent $i$ goes through the same sequence of information sets before reaching the two histories. If the divergence results from agent $j$'s decision in a renunciation or an assertion sub-stage, it is clear that agent $j$ must have left the market before $h^1$ or $h^2$.

Suppose instead that agent $j$ initiates the divergence in a designation sub-stage of stage $t$. By the information structure of RDA mechanisms, agent $j$'s different designations are unobservable to any other agent in stage $t$. Since agents other than $i$ and $j$ follow the common strategy profile $s_{-i,j}$ and agent $i$ behaves identically before $h^1$ and $h^2$, all agents other than $j$ act identically on the two paths in stage $t$. If agent $j$ does not enter a trading cycle and leave the market on either path, then the same trading cycle forms in stage $t$ on both paths. This implies that no agent other than $j$ can distinguish the two paths in stage $t+1$, so they must act identically in the new stage. Applying this line of reasoning repeatedly in a mathematical induction, agent $i$ can distinguish $h^1$ and $h^2$ only if agent $j$ has left the market before $h^1$ or $h^2$. The formal proof is in Appendix \ref{proofsec4}.

\section{Conclusion}
\label{seccon}

For any dynamic mechanism implementing an SCF in dominant type-strategies, one can construct an incentive compatible gradual mechanism implementing the same SCF by pruning off-equilibrium-path histories and relabeling each remaining action with a set of types that choose the action in equilibrium \citep[an operation appearing repeatedly in the literature, see][]{Li2017, Mackenzie2020, CW2024}. The advantages of dynamic mechanism implementations discussed in the literature usually arise from the information transmission on the equilibrium path \citep[e.g.,][]{Akbarpour2020,haupt2024}. In this sense, the class of gradual mechanisms may provide a minimal but sufficient domain to characterize dominant strategy implementation for a designer whose aim is to harness the benefits of dynamic mechanisms.

In this paper, we offer two characterizations of when truth-telling remains a dominant strategy in gradual mechanisms as they sequentialize the one-shot information revelation in direct mechanisms of strategy-proof SCFs. The first, based on three transformations that can reduce any gradual mechanism into the direct mechanism, hinges on the incentive-preservation condition for the illuminating transformation that partitions one agent's information sets. The second builds on the reaction-proofness condition for gradual mechanisms, which requires that an agent's reaction to any other agent, i.e., reporting differently immediately after acquiring some information, does not harm the other agent's incentive for truth-telling. The characterizations are applied to the design of dynamic mechanisms for the second-price auction rule and the hierarchical exchange rule. In the first application, we define a notion of transparency for gradual mechanisms in which any further information transmission will break incentive compatibility, operationalized by the absence of opportunities for incentive-preserving illuminating, and present a transparent ascending-price auction. In the second application, we use our reaction-proofness condition to establish the incentive compatibility of renunciation-designation-assertion mechanisms \citep{CW2024}, which extend the dynamic mechanisms in \cite{Troyan2019} and \cite{mandal2022obviously} to environments where obviously dominant strategy implementations are unavailable.

\appendix
\clearpage

{\centering\LARGE{Appendices}\par}
\vspace*{0.5cm}
\section{The Definition of Dynamic Game Forms}
\label{sec:app_definitions}

\noindent In a dynamic game form $G = (\bar{H}, \{A_i, \boldsymbol{H}_i\}_{i\in N}, \mathcal{X})$ with perfect recall, there are:

\begin{itemize}
    \item \emph{Players.} $N$ is finite and each $i\in N$ is a player.
    \item \emph{Actions.} For each player $i\in N$, $A_i$ is a non-empty set of her available actions. Denote the set of action profiles by \[A = \bigcup_{M\subseteq N} \prod_{i\in M} A_i.\]
    \begin{itemize}
       \item For each player $i \in N$, let $A_{-i}=\bigcup_{M \subseteq N \setminus \{i\}}(\prod_{j \in M} A_j)$ be the set of action profiles in which player $i$ is not active. For any $a_i \in A_i$ and $a_{-i} \in A_{-i}$, let $(a_i,a_{-i}) \in A$ denote the action profile defined by their combination. Specifically, let $(a_i,a_{-i})$ be $(a_i)$ if $a_{-i}=\varnothing$.
       \item For each integer $T > 0$, let $A^T$ denote the collection of sequences of action profiles of length $T$, with a generic element being denoted by $h = \langle h^{(1)}, \ldots, h^{(T)} \rangle$. Let $A^0=\{\varnothing\}$ and $A^{<\mathbb{N}} = \bigcup_{T\in \mathbb{N}} A^T$ where $\mathbb{N}=\{0,1,...\}$.
       \item Define a precedence relation $\preceq$ on $A^{<\mathbb{N}}$: for any $h \in A^S$ and any $\bar{h} \in A^T$, $h \preceq \bar{h}$ if $S=0$ or if $0 < S \leq T$ with $h^{(s)}=\bar{h}^{(s)}$ for any $1 \leq s \leq S$. Let $\prec$ be the asymmetric part of $\preceq$, i.e., $h \prec \bar{h}$ when $h \preceq \bar{h}$ and $h \neq \bar{h}$. If $h \prec \bar{h}$ with $h \in A^T$ and $\bar{h} \in A^{T+1}$, say $h$ is an immediate predecessor of $\bar{h}$. Note that any non-empty $h \in A^{<\mathbb{N}}$ has a unique immediate predecessor.
       
       \item For any $h,g \in A^{\mathbb{N}}$ with $h=\langle h^{(1)},...,h^{(k)} \rangle$ and $g=\langle g^{(1)},...,g^{(m)} \rangle$, define $h \oplus g=\langle h^{(1)},...,h^{(k)},g^{(1)},...,g^{(m)} \rangle$. In this expression, an element $\langle a \rangle \in A^1$ and the action profile $a \in A$ are used interchangeably. In particular, let $\varnothing \oplus h = h \oplus \varnothing = h$ for any $h \in A^{<\mathbb{N}}$.

    \end{itemize}
    \item \emph{Histories.} The set of histories $\bar{H}$ is a finite tree contained in $(A^{<\mathbb{N}_0},\preceq)$ such that $\varnothing \in \bar{H}$ and for any history $h \in \bar{H}$ with $h \neq \varnothing$, its immediate predecessor is in $\bar{H}$.
\begin{itemize}
    \item Denote the set of terminal histories ($\preceq$-maximal elements of $\bar{H}$) by $Z$ and non-terminal histories by $H$. For any $h \in H$, let $\sigma(h)$ denote the collection of its immediate successors in $\bar{H}$.

    \item We can derive a non-empty valued active-player correspondence $\mathbb{P} : H \twoheadrightarrow N$, capturing the set of players that are simultaneously active at a particular non-terminal history, such that for any $h \in H$ and any $a \in A$ satisfying $h \oplus a \in \sigma(h)$, it is the case that $a \in \prod_{i \in \mathbb{P}(h)} A_i$.
    
    \item Let $H_i = \{h \in H : i \in \mathbb{P}(h)\}$ represent the collection of non-terminal histories at which player $i$ is active, whose elements are also referred to as her decision nodes. For each $h \in H_i$, define $A_i(h) = \{a_i \in A_i : h \oplus (a_i, a_{-i}) \in \bar{H} \text{ for some } a_{-i} \in A_{-i}\}$, which is the collection of available actions of player $i$ at this decision node.
    
    \item Assume that for any history $h \in H$ and any action profile $a \in \prod_{i \in \mathbb{P}(h)} A_i(h)$, the history $h \oplus a$ belongs to $\bar{H}$.
    
    \item Assume $\lvert A_i(h) \rvert \geq 2$ for any $\varnothing \neq h \in H$ and any $i \in \mathbb{P}(h)$. We allow the initial history $\varnothing$ to be a \textit{degenerate} decision node for some player $i$, i.e., $A_i(\varnothing)$ is a singleton. Furthermore, assume that every player is active at the initial history and $\lvert A_i(\varnothing) \rvert \geq 2$ for some $i \in N$. 
\end{itemize}

\item \emph{Outcomes.} $\mathcal{X}:Z\rightarrow X$ assigns each terminal history a public outcome.

    \item \emph{Information Sets.} For each $i\in N$, $\boldsymbol{H}_i$ is a partition of $H_i$ whose elements are called information sets, such that:
    \begin{itemize}
        \item For each player $i \in N$, $A_i$ is $\boldsymbol{H}_i$-measurable, i.e., for any $\boldsymbol{h}_i\in \boldsymbol{H}_i$ and any $h, h'\in \boldsymbol{h}_i$, we have $A_i(h) = A_i(h')$. Let $A(\boldsymbol{h}_i) = A_i(h)$ with $h\in \boldsymbol{h}_i$ denote the available actions for player $i$ at $\boldsymbol{h}_i$.
        \item The game form $G$ has perfect recall. That is, before any two histories in an information set, player $i$ has encountered the same sequence of information sets at which she has taken the same actions. See \cite{osborne1994course}'s Definition $203.3$.  
        
    \end{itemize}
   
\end{itemize}

Note that given perfect recall and $\mathbb{P}(\varnothing)=N$, the singleton set of the initial history $\{\varnothing\}$ is the initial information set for each player. Under perfect recall, a precedence relation on $\boldsymbol{H}_i$ can be defined: $\boldsymbol{h}_i\preceq \bar{\boldsymbol{h}}_i$ ($\boldsymbol{h}_i\prec \bar{\boldsymbol{h}}_i$) if there exist $h\in \boldsymbol{h}_i$ and $\bar{h}\in \bar{\boldsymbol{h}}_i$ such that $h\preceq \bar{h}$ ($h\prec \bar{h}$). Say $\bar{\boldsymbol{h}}_i$ is an immediate successor of $\boldsymbol{h}_i$ if $\boldsymbol{h}_i \prec \bar{\boldsymbol{h}}_i$ and there is no $\tilde{\boldsymbol{h}}_i$ such that $\boldsymbol{h}_i \prec \tilde{\boldsymbol{h}}_i \prec \bar{\boldsymbol{h}}_i$. Denote the set of immediate successors of $\boldsymbol{h}_i$ by $\sigma(\boldsymbol{h}_i)$ and its subset consistent with player $i$ choosing $a_i$ at $\boldsymbol{h}_i$ by $\sigma_{a_i}(\boldsymbol{h}_i)$. The precedence relation can also be extended to between an information set $\boldsymbol{h}_i$ and a history $\bar{h}$: $\boldsymbol{h}_i \preceq \bar{h}$ ($\boldsymbol{h}_i \prec \bar{h}$) if there exists $h\in \boldsymbol{h}_i$ such that $h\preceq\bar{h}$ ($h\prec\bar{h}$).

For any player $i\in N$, a pure strategy $s_i: H_i\rightarrow A_i$ specifies an available action $s_i(h)\in A_i(h)$ for each history $h\in H_i$ such that $s_i(h) = s_i(h')$ if $h$ and $h'$ belong to the same information set. Therefore, for an information set $\boldsymbol{h}_i$, we sometimes abuse notation and write $s_i(\boldsymbol{h}_i) = s_i(h)$ for any $h \in \boldsymbol{h}_i$. We use $S_i$ to denote the set of strategies for each player $i \in N$.

\section{Three Basic Transformations}
\label{sec:app_transformations}
\subsectionfont{\normalfont\fontsize{12}{13}\selectfont}

\subsection{Splitting}
Let $\boldsymbol{h}_i$ be an information set of some agent $i$ in a gradual mechanism $G$ with $a_i \in A(\boldsymbol{h}_i)$. Suppose $\bar{\boldsymbol{h}}_i=\{z \in Z: \Theta_i(z)=a_i \mbox{ and } \boldsymbol{h}_i\prec z\}$ is not empty. Let $\{a_i^1, a_i^2\}$ be a partition of $a_i$. Then, SPL requires agent $i$ to additionally choose between $a_i^1$ and $a_i^2$ at the new information set $\bar{\boldsymbol{h}}_i$. Formally, it delivers a gradual mechanism $G^*$ in the following manner:

\begin{itemize}
  \item The set of histories in $G^*$ is given by $\bar{H}^*=\bar{H} \cup \{z \oplus a^k_i\}_{z \in \bar{\boldsymbol{h}}_i, a^k_i \in \{a^1_i,a^2_i\}}$.
  \item Agent $j$'s (for each $j\neq i$) information sets in $G^*$ are invariant, i.e., $\boldsymbol{H}^*_j=\boldsymbol{H}_j$.
  \item For agent $i$, we have $\boldsymbol{H}_{i}^{*}=\boldsymbol{H}_i \cup \{\bar{\boldsymbol{h}}_i\}$.
  \item There exists an onto mapping $T: Z^*\to Z$ by $\Theta(z^*) \subseteq \Theta(T(z^*))$ for each $z^* \in Z^*$. The outcome function $\mathcal{X}^*$ in $G^{*}$ is given by $\mathcal{X}^*(z^*) = \mathcal{X}(T(z^*))$.
\end{itemize}

\subsection{Coalescing}

There is a COA opportunity in a gradual mechanism $G$ if there exist two information sets $\boldsymbol{h}^c_i$ and $\bar{\boldsymbol{h}}^c_i$ of some agent $i$ and an available action $a_i\in A(\boldsymbol{h}^c_i)$ such that $\bar{\boldsymbol{h}}^c_i\in \sigma_{a_i}(\boldsymbol{h}^c_i)$ and $\Theta_{-i}(\boldsymbol{h}^c_i) = \Theta_{-i}(\bar{\boldsymbol{h}}^c_i)$. Formally, it delivers a gradual mechanism $G^*$ in the following manner:
\begin{itemize}
    \item Define histories in $G^*$ and a mapping $T: \bar{H}^* \to \bar{H}$ in the following ways:
    \begin{enumerate}
        \item For any history $h \in \bar{H}$ such that for any $h^c\in \boldsymbol{h}^c_i$ and any $a_{-i} \in A_{-i}$, it is not $h^c \oplus (a_i, a_{-i})\preceq h$: let $h \in \bar{H}^*$ and $T(h)=h$.
        \item For any history $h \in \bar{H}$ such that (i) $h=h^c \oplus (a_i,a_{-i}) \oplus h'$ for some $h^c \in \boldsymbol{h}^c_i$, $a_{-i} \in A_{-i}$ and $h' \in A^{< \mathbb{N}}$, (ii) for any $\bar{h}^c \in \bar{\boldsymbol{h}}^c_i$, it is not $\bar{h}^c \prec h$, and (iii) $h \in \bar{\boldsymbol{h}}^c_i$ implies $\mathbb{P}(h)\neq \{i\}$: for each $\bar{a}_i \in A(\bar{\boldsymbol{h}}^c_i)$, let $h^*=h^c \oplus (\bar{a}_i,a_{-i}) \oplus h'$ be in $\bar{H}^*$ and $T(h^*)=h$.
        \item For any history $h \in \bar{H}$ such that $h=\bar{h}^c \oplus (\bar{a}_i,\bar{a}_{-i}) \oplus h'$ for some $\bar{h}^c \in \bar{\boldsymbol{h}}^c_i$, $\bar{a}_i \in A(\bar{\boldsymbol{h}}^c_i)$, $\bar{a}_{-i} \in A_{-i}$, and $h' \in A^{<\mathbb{N}}$, where $\bar{h}^c=h^c \oplus (a_i,a_{-i}) \oplus g$ for some $h^c \in \boldsymbol{h}^c_i$, $a_{-i} \in A_{-i}$, and $g \in A^{<\mathbb{N}}$: let $h^*=h^c \oplus (\bar{a}_i,a_{-i}) \oplus g \oplus \bar{a}_{-i} \oplus h'$ be in $\bar{H}^*$ and $T(h^*)=h$.
    \end{enumerate}
For each information set $\boldsymbol{h}_j \in \boldsymbol{H}_j$ of each agent $j \in N$, let $T^{-1}(\boldsymbol{h}_j) = \{h^*\in H_j^*: T(h^*)\in \boldsymbol{h}_j\}$. Note that only $T^{-1}(\bar{\boldsymbol{h}}^c_i) = \varnothing$.
    
    \item For each agent $j \in N$, the collection of information sets in $G^*$ is given by $\boldsymbol{H}_j^* = \{T^{-1}(\boldsymbol{h}_j): \boldsymbol{h}_j\in\boldsymbol{H}_j \mbox{ such that } T^{-1}(\boldsymbol{h}_j)\neq \varnothing\}$.  
    \item The outcome function $\mathcal{X}^*$ in $G^*$ is given by $\mathcal{X}^*(z^*) = \mathcal{X}(T(z^*))$.
\end{itemize}

\subsection{Illuminating}

An ILL transformation partitions an information set $\boldsymbol{h}_i$ of some agent $i$ in a gradual mechanism $G$ into two non-empty information sets $\boldsymbol{h}^{1}_i$ and $\boldsymbol{h}^{2}_i$. To preserve the perfect recall assumption, it also partitions each successive information set of $\boldsymbol{h}_i$ accordingly. For each $\bar{\boldsymbol{h}}_i$ with $\boldsymbol{h}_i\preceq \bar{\boldsymbol{h}}_i$, define $\bar{\boldsymbol{h}}^{k}_i=\{\bar{h} \in \bar{\boldsymbol{h}}_i: \boldsymbol{h}^k_i\preceq \bar{h}\}$ for both $k = 1,2$. Note that some $\bar{\boldsymbol{h}}^{k}_i$ thus defined may be empty. Formally, ILL delivers a gradual mechanism $G^*$ in the following manner:

\begin{itemize}
  \item The collection of histories in $G^*$ is invariant, i.e., $\bar{H}^* = \bar{H}$.
  \item Agent $j$'s (for each $j\neq i$) information sets in $G^*$ are invariant, i.e., $\boldsymbol{H}^*_j=\boldsymbol{H}_j$.
  \item Agent $i$'s information sets are given by $\boldsymbol{H}_{i}^*=\{\hat{\boldsymbol{h}}_i \in \boldsymbol{H}_{i}: \mbox{not }\boldsymbol{h}_{i}\preceq \hat{\boldsymbol{h}}_i\} \cup \{\bar{\boldsymbol{h}}_{i}^{k}: \bar{\boldsymbol{h}}_i \in \boldsymbol{H}_i \mbox{ and } k\in\{1,2\} \mbox{ such that } \boldsymbol{h}_{i}\preceq \bar{\boldsymbol{h}}_i \mbox{ and } \bar{\boldsymbol{h}}_{i}^{k} \neq \varnothing\}$.
  \item The outcome function in $G^{*}$ is invariant, i.e., $\mathcal{X}^*=\mathcal{X}$.
\end{itemize}

\section{Proofs}

\subsection{Proofs for Section \ref{secframe}}
\label{proofsec2}
\begin{proof}[Proof of Proposition \ref{propnec}]
  If SCF $f$ is not strategy-proof, there exist $\theta_i, \theta'_i \in \Theta_i$ and $\theta_{-i} \in \Theta_{-i}$ such that $f(\theta_i,\theta_{-i})P(\theta'_i)f(\theta'_i,\theta_{-i})$. For any gradual mechanism $G$ implementing $f$, take the unconditional strategies $s_{\theta_i},s_{\theta'_i} \in S_i$ and $s_{\theta_{-i}}=(s_{\theta_j})_{j \in N \setminus \{i\}} \in S_{-i}$ for those types. As $f(\theta_i,\theta_{-i})=\mathcal{X}(s_{\theta_i},s_{\theta_{-i}})P(\theta'_i)\mathcal{X}(s_{\theta'_i},s_{\theta_{-i}})=f(\theta'_i,\theta_{-i})$, $G$ is not incentive compatible. 
\end{proof}

\begin{lemma}
  \label{lemmaunconditional}
  Let $h \in \bar{H}$ be a history that is consistent with a strategy profile $s_M \in S_M$ for some $\varnothing \neq M \subseteq N$ in a gradual mechanism $G$. 
  
  (i) For any $i \in M$ and $s'_i \in S_i$ such that $s'_i(\boldsymbol{h}_i)=s_i(\boldsymbol{h}_i)$ for any $\boldsymbol{h}_i \in \boldsymbol{H}_i$ with $\boldsymbol{h}_i \prec h$, $(s'_i,(s_j)_{j \in M, j \neq i})$ is consistent with $h$.

  (ii) In particular, for any $\theta \in \Theta(h)$ and any $i \in M$, $(s_{\theta_i},(s_j)_{j \in M, j \neq i})$ is consistent with $h$, where $s_{\theta_i}$ is an unconditional strategy for $\theta_i$.
\end{lemma}

\begin{proof}
  The first part of Lemma \ref{lemmaunconditional} is straightforward, and we prove the second part. Let $h \in \bar{H}$ be on the path determined by $(s_{N \setminus M},s_M) \in S$ for some $\varnothing \neq M \subseteq N$ and $s_M \in S_M$. Take any $i \in M$, $\theta\in \Theta(h)$, and any unconditional strategy $s_{\theta_i}$ for $\theta_i$. Then, for any agent $i$'s decision node $\underline{h} \in H_i$ with $\underline{h} \prec h$, there is a unique action $a_i \in A_i(\underline{h})$ such that $\theta_i \in a_i$ and $s_i(\underline{h})=s_{\theta_i}(\underline{h})=a_i$. Therefore, $h$ is also on the path determined by $(s_{N \setminus M},s_{\theta_i},(s_j)_{j \in M, j \neq i})$.
\end{proof}

\begin{proof}[Proof of Proposition \ref{Unconditional2}]
  \textit{``If'' part.} Arbitrarily take $i \in N$, $\theta_i \in \Theta_i$, unconditional strategy $s_{\theta_i}$ for $\theta_i$, $s_i \in S_i$, and $s_{-i} \in S_{-i}$. Take any $\theta^2 \in \Theta(z(s_i,s_{-i}))$ and any $\theta^1 \in \Theta(z(s_{\theta_i},s_{-i}))$ with $\theta^1_i=\theta_i$. As $\theta^1$ and $\theta^2$ are consistent with $s_{-i}$, by the assumption in the ``if'' direction, we have $\mathcal{X}(s_{\theta_i},s_{-i})=f(\theta^1)R(\theta_i)f(\theta^2)=\mathcal{X}(s_i,s_{-i})$.

  \textit{``Only if'' part.} Take any $i \in N$ and $\theta^1,\theta^2 \in \Theta$ that are consistent with a common $s_{-i} \in S_{-i}$. Let $s^1_i, s^2_i \in S_i$ be such that $\theta^1 \in \Theta(z(s^1_i,s_{-i}))$ and $\theta^2 \in \Theta(z(s^2_i,s_{-i}))$. By Lemma \ref{lemmaunconditional}, $z(s^1_i,s_{-i})=z(s_{\theta^1_i},s_{-i})$, where $s_{\theta^1_i}$ is an unconditional strategy for $\theta^1_i$. Incentive compatibility implies that $f(\theta^1)=\mathcal{X}(s_{\theta^1_i},s_{-i})R(\theta^1_i)\mathcal{X}(s^2_i,s_{-i})=f(\theta^2)$.
\end{proof}

\subsection{Proofs for Section \ref{secchara}}
\label{proofsec3}

The following Lemma \ref{lemmaunconditional2} is an immediate consequence of Lemma \ref{lemmaunconditional}, and we omit the proof. It suggests that, to verify the incentive compatibility of a gradual mechanism, we only need to check the dominance of truth-telling among the unconditional strategies for each agent.

\begin{lemma}
  \label{lemmaunconditional2}
  A gradual mechanism $G$ implementing an SCF $f$ is incentive compatible if and only if for any agent $i \in N$ and any $\theta_i^1, \theta_i^2\in \Theta_i$, it is the case that $\mathcal{X}(s_{\theta_i^1}, s_{-i}) R(\theta_i^1) \mathcal{X}(s_{\theta_i^2}, s_{-i})$ for any $s_{-i}\in S_{-i}$ where $s_{\theta_i^1}$ and $s_{\theta_i^2}$ are unconditional strategies for $\theta_i^1$ and $\theta_i^2$, respectively.
\end{lemma}

In the following proofs of propositions \ref{propsplic}-\ref{propillic}, we observe how the sets $S_i$ and $S_i^*$ of agent $i$'s strategies relate to each other as a basic transformation concerning agent $i$ changes a gradual mechanism $G$ into another $G^*$. Such relations are the key to the proofs since the strategies $S_j$ and $S_j^*$ of any other agent $j$ are essentially invariant. Lemmas \ref{lemmaunconditional} and \ref{lemmaunconditional2} will be used throughout without mentioning.

\begin{proof}[Proof of Proposition \ref{propsplic}]
Suppose a gradual mechanism $G$ can be transformed into another $G^*$ through an SPL identifiable by $\boldsymbol{h}_i \in \boldsymbol{H}_i$, $a_i \in A(\boldsymbol{h}_i)$, $a^1_i \cup a^2_i=a_i$, and $\bar{\boldsymbol{h}}_i \subseteq Z$. For each agent $j \neq i$, we have $S_j^* = S_j$ since each agent $j$ has the same collection of information sets in $G$ and $G^*$. For agent $i$, $S^*_i$ is derived from $S_i$ as follows: for each $s_i\in S_i$, there are two strategies, denoted by $s_{a_i^1}$ and $s_{a_i^2}$, in $S^*_i$, such that $s_{a_i^1}(\boldsymbol{h}_i) = s_{a_i^2}(\boldsymbol{h}_i) = s(\boldsymbol{h}_i)$ for each $\boldsymbol{h}_i\in \boldsymbol{H}_i$ while $s_{a_i^1}(\bar{\boldsymbol{h}}_i) = a_i^1$ and $s_{a_i^2}(\bar{\boldsymbol{h}}_i) = a_i^2$. Then, it is the case that $\mathcal{X}(s_i, s_{-i}) = \mathcal{X}^*(s_{a_i^1}, s_{-i}) = \mathcal{X}^*(s_{a_i^2}, s_{-i})$ for each $s_i\in S_i$ and each $s_{-i}\in S_{-i}$. Next, we only show that if $G$ is incentive compatible, then truth-telling is a dominant strategy for agent $i$ in $G^*$.

If $s_i$ is an unconditional strategy for some $\theta_i\not\in a_i$ in $G$, both $s_{a_i^1}$ and $s_{a_i^2}$ are unconditional strategies for $\theta_i$ in $G^*$. If $s_i$ is an unconditional strategy for some $\theta_i\in a_i$ in $G$, there exists $k\in \{1, 2\}$ such that $\theta_i\in a_i^k$. Therefore, for each $\theta_i\in \Theta_i$ and each unconditional strategy $s_i$ for $\theta_i$ in $G$, there exists $s_{a_i^k}$ that is an unconditional strategy for $\theta_i$ in $G^*$. Suppose $G$ is incentive compatible. To show $\mathcal{X}^*(s_{a_i^k}, s_{-i}) R(\theta_i) \mathcal{X}^*(s_i^*, s_{-i})$ for each $s_i^*\in S_i^*$ and each $s_{-i}\in S_{-i}$, notice that (i) $\mathcal{X}^*(s_{a_i^k}, s_{-i}) = \mathcal{X}(s_i, s_{-i})$, (ii) there exists $s_i'\in S_i$ such that $\mathcal{X}^*(s_i^*, s_{-i}) = \mathcal{X}(s_i', s_{-i})$, and (iii) $s_i$ is unconditional for $\theta_i$ in $G$.
\end{proof}

\begin{proof}[Proof of Proposition  \ref{propcoaic}]
Suppose a gradual mechanism $G$ can be transformed into another $G^*$ by a COA identifiable by $\boldsymbol{h}_i^c, \bar{\boldsymbol{h}}_i^c \in \boldsymbol{H}_i$ and $a_i \in A(\boldsymbol{h}_i^c)$. By the definition of COA, there exist:

\begin{enumerate}
  \item a bijection $T: Z \to Z^*$ such that $\Theta(z)=\Theta(T(z))$ for any $z \in Z$,
  \item for each agent $j\neq i$, a bijection $I_j: \boldsymbol{H}_j \to \boldsymbol{H}^*_j$ such that $\Theta(\boldsymbol{h}_j)=\Theta(I_j(\boldsymbol{h}_j))$ and $A(\boldsymbol{h}_j)=A(I_j(\boldsymbol{h}_j))$ for any $\boldsymbol{h}_j \in \boldsymbol{H}_j$,
  \item a bijection $I_i: \boldsymbol{H}_i \setminus \{\bar{\boldsymbol{h}}_i^c\} \to \boldsymbol{H}^*_i$ such that (i) $\Theta(\boldsymbol{h}_i)=\Theta(I_i(\boldsymbol{h}_i))$ for any $\boldsymbol{h}_i \in \boldsymbol{H}_i \setminus \{\bar{\boldsymbol{h}}^c_i\}$, (ii) $A(\boldsymbol{h}_i)=A(I_i(\boldsymbol{h}_i))$ for any $\boldsymbol{h}_i \in \boldsymbol{H}_i \setminus \{\boldsymbol{h}_i^c,\bar{\boldsymbol{h}}^c_i\}$, and (iii) $A(I_i(\boldsymbol{h}^c_i))=A(\boldsymbol{h}^c_i) \cup A(\bar{\boldsymbol{h}}^c_i) \setminus \{a_i\}$.
\end{enumerate}

Given the above bijections between terminal histories and agents' information sets in $G$ and $G^*$, we can also define correspondences between agents' strategies. For each agent $j \neq i$, there exists a bijection $g_j: S_j \to S^*_j$ such that $s_j(\boldsymbol{h}_j)=g_j(s_j)(I_j(\boldsymbol{h}_j))$ for any $\boldsymbol{h}_j \in \boldsymbol{H}_j$ and any $s_j \in S_j$. For agent $i$, there is an onto mapping $g_i: S_i \to S^*_i$ such that for any $s_i \in S_i$, (i) $s_i(\boldsymbol{h}_i)=g_i(s_i)(I_i(\boldsymbol{h}_i))$ for any $\boldsymbol{h}_i \in \boldsymbol{H}_i \setminus \{\boldsymbol{h}_i^c,\bar{\boldsymbol{h}}^c_i\}$, (ii) $s_i(\boldsymbol{h}^c_i)=g_i(s_i)(I_i(\boldsymbol{h}^c_i))$ if $s_i(\boldsymbol{h}^c_i) \neq a_i$, and (iii) $s_i(\bar{\boldsymbol{h}}^c_i)=g_i(s_i)(I_i(\boldsymbol{h}^c_i))$ if $s_i(\boldsymbol{h}^c_i) = a_i$. Now, we have $T(z(s))=z^*(g(s))$ for any $s \in S$ where $g(s)=(g_j(s_j))_{j \in N}$. Also, notice that any $s_j \in S_j$ for any $j \in N$ is unconditional in $G$ if and only if $g_j(s_j)$ is unconditional in $G^*$. The rest is straightforward.
\end{proof}

\begin{proof}[Proof of Proposition \ref{propillic}]

  Suppose a gradual mechanism $G$ can be transformed into another $G^*$ through an ILL that splits $\boldsymbol{h}_i$ into $\boldsymbol{h}_i^1$ and $\boldsymbol{h}_i^2$. For each agent $j \neq i$, we have $S_j^* = S_j$ since each agent $j$ has the same collection of information sets in $G$ and $G^*$. For agent $i$, each strategy $s^*_i \in S_i^*$ can be identified by a pair $(s_i^1, s_i^2) \in S_i\times S_i$, such that $s_i^*$ behaves as $s_i^2$ at any information set $\bar{\boldsymbol{h}}_i^2$, as defined in Appendix B.3, for each $\bar{\boldsymbol{h}}_i$ with $\boldsymbol{h}_i\preceq \bar{\boldsymbol{h}}_i$, i.e., $s_i^*(\bar{\boldsymbol{h}}_i^2) = s_i^2(\bar{\boldsymbol{h}}_i)$, and otherwise behaves as $s_i^1$.\footnote{The construction of $s_i^*$ exploits a feature of game forms with perfect recall named strategic independence \citep{mailath1993extensive}.} Any unconditional strategy $s_{\theta_i}^*$ for $\theta_i$ in $G^*$ can be identified by $(s_{\theta_i}, s_{\theta_i}')$ where $s_{\theta_i}$ and $s_{\theta_i}'$ are both unconditional strategies for $\theta_i$ in $G$ behaviorally equivalent to each other. Therefore, for any ILL, truth-telling is a dominant strategy for agent $i$ in $G^*$ if and only if it is so in $G$. It follows directly that $G$ is incentive compatible when $G^*$ is incentive compatible. Next, we consider the truth-telling incentives for agents other than $i$ in $G^*$ when $G$ is incentive compatible.

Take any agent $j \neq i$ and any $\theta^1_j,\theta^2_j \in \Theta_j$. Let $s_{\theta^1_j}$ and $s_{\theta^2_j}$ be the unconditional strategies for these two types. Take any $s_i^* \in S_i^* \setminus S_i$ identified by $(s_i^1, s_i^2) \in S_i \times S_i$ and any $s_{-i,j}\in S_{-i,j} = S_{-i,j}^*$. Denote $z^1 = z(s_{\theta^1_j}, s_i^*, s_{-i,j}) \in Z^*$ and $z^2 = z(s_{\theta^2_j}, s_i^*, s_{-i,j}) \in Z^*$. Without loss of generality, consider the following two cases.
 
 \begin{enumerate}
  \item Suppose $\boldsymbol{h}^1_i\prec z^1$ and $\boldsymbol{h}^2_i\prec z^2$. Then, any pair of $(\theta_i^1, \theta_j^1,\theta^1_{-i,j}) \in \Theta(z^1)$ and $(\theta_i^2, \theta_j^2,\theta^2_{-i,j}) \in \Theta(z^2)$ satisfies (i) $(\theta_j^1, \theta^1_{-i,j})\in \Theta_{-i}(\boldsymbol{h}_i^1)$, (ii) $(\theta_j^2, \theta^2_{-i,j})\in \Theta_{-i}(\boldsymbol{h}_i^2)$, (iii) $(\theta_i^1, \theta_j^1,\theta^1_{-i,j})$ and $(\theta_i^2, \theta_j^2,\theta^2_{-i,j})$ are both consistent with $s_{-i,j}$. In this case, $\mathcal{X}^*(z^1)R(\theta^1_j)\mathcal{X}^*(z^2)$ if and only if the ILL is incentive-preserving.
  \item Suppose neither $\boldsymbol{h}^2_i\prec z^1$ nor $\boldsymbol{h}^2_i\prec z^2$. Then, $\mathcal{X}^*(z^1) = \mathcal{X}(s_{\theta^1_j}, s_i^1, s_{-i,j})$ and $\mathcal{X}^*(z^2) = \mathcal{X}(s_{\theta^2_j}, s_i^1, s_{-i,j})$. In this case, $\mathcal{X}^*(z^1)R(\theta^1_j)\mathcal{X}^*(z^2)$ if $G$ is incentive compatible.
 \end{enumerate}

The above analysis shows that if $G$ is incentive compatible, then $G^*$ is incentive compatible if and only if the ILL is incentive-preserving, completing the proof. 
\end{proof}

The following lemmas \ref{lemmanoSPL}-\ref{lemmastatic} are used to prove Proposition \ref{Reduction}. 

\begin{lemma}
  If a gradual mechanism $G$ has no SPL opportunity and $G'$ is transformed from $G$ through a COA or an inverse ILL, then $G'$ has no SPL opportunity.
  \label{lemmanoSPL}
\end{lemma}

\begin{proof}
  It is obvious that a gradual mechanism $G$ has no SPL opportunity if and only if $\Theta(z)$ is a singleton for each terminal history $z\in Z$. Neither a COA nor an inverse ILL changes the accrued information at terminal histories. Therefore, the lemma holds.
\end{proof}

\begin{lemma}
  If a non-direct gradual mechanism $G$ has no SPL opportunity, then it must have a COA opportunity or an inverse ILL opportunity.
  \label{lemmastatic}
\end{lemma}
\begin{proof}
If a gradual mechanism $G$ is not a direct mechanism, some agent $i \in N$ has at least two information sets. Let $\boldsymbol{h}_i \in \boldsymbol{H}_i$ be a $\preceq$-maximal information set of agent $i$, and let $\underline{\boldsymbol{h}}_i$ denote its immediate predecessor. If $G$ has no SPL opportunity, then there is no $z \in Z$ such that $\underline{\boldsymbol{h}}_i \prec z$ and $\Theta_i(z)=\Theta_i(\boldsymbol{h}_i)$. Suppose further that $G$ has no inverse ILL opportunity. There are two cases.

(i) $\boldsymbol{h}_i$ is the unique immediate successor of $\underline{\boldsymbol{h}}_i$ following the action $\Theta_i(\boldsymbol{h}_i)$. In this case, the two consecutive information sets $\underline{\boldsymbol{h}}_i$ and $\boldsymbol{h}_i$ satisfy $\Theta_{-i}(\underline{\boldsymbol{h}}_i)=\Theta_{-i}(\boldsymbol{h}_i)$, implying a COA opportunity.

(ii) There exists another immediate successor $\boldsymbol{h}'_i$ of $\underline{\boldsymbol{h}}_i$ such that $\Theta_i(\boldsymbol{h}'_i)=\Theta_i(\boldsymbol{h}_i)$ and $A(\boldsymbol{h}'_i)\neq A(\boldsymbol{h}_i)$. In this case, there is a $\preceq$-maximal information set $\bar{\boldsymbol{h}}'_i$ following $\boldsymbol{h}'_i$, to which the same reasoning applies.

Since $G$ is finite, the iteration of case (ii) must eventually terminate in case (i), and hence $G$ admits a COA opportunity.
\end{proof}

\begin{proof}[Proof of Proposition \ref{Reduction}]
Suppose a gradual mechanism $G^1$ can be transformed into $G^2$ by a COA or an inverse ILL with respect to agent $i$. Then, observe that the number of agent $i$'s information sets in $G^2$ is less than that in $G^1$.

 Let $G$ be an arbitrary non-direct gradual mechanism. It is straightforward to apply a sequence of SPLs to $G$ to obtain a mechanism $G'$ in which no further SPL is applicable. Then, by lemmas \ref{lemmanoSPL} and \ref{lemmastatic}, and the previous observation, there is a sequence of COAs and inverse ILLs that transforms $G'$ into another $G^*$ in which each agent $i \in N$ has a unique information set $\boldsymbol{h}_i=\{\varnothing\}$ with $A(\boldsymbol{h}_i)=\Theta_i$. Thus, $G^*$ is a direct mechanism.
\end{proof}

\begin{proof}[Proof of Proposition \ref{propscf}]
Let $f$ be a strategy-proof SCF such that for any agent $i \in N$, any distinct $\theta^1_i,\theta^2_i \in \Theta_i$, and any non-trivial partition $\{R^1_{-i},R^2_{-i}\}$ of $\Theta_{-i}$, there exist $j \in N$ with $j\neq i$, $(\theta^1_j,\theta_{-i,j}) \in R^1_{-i}$ and $(\theta^2_j,\theta_{-i,j}) \in R^2_{-i}$ such that for some $k,l \in \{1,2\}$, it is the case that $f(\theta^k_j,\theta^l_i,\theta_{-i,j})P(\theta^{3-k}_j)f(\theta^{3-k}_j,\theta^{3-l}_i,\theta_{-i,j})$.\footnote{Since $\Theta_{-i}$ is a Cartesian product, the associated graph is connected. Hence, for any non-trivial partition $\{R^1_{-i},R^2_{-i}\}$, there exist $\theta^1_{-i} \in R^1_{-i}$ and $\theta^2_{-i} \in R^2_{-i}$ that differ in exactly one coordinate.} Let $G$ be a non-static gradual mechanism implementing $f$ and $\boldsymbol{h}^*_i$ be an information set of some agent $i$ with $\Theta_{-i}(\boldsymbol{h}^*_i) \subsetneq \Theta_{-i}$.

\textit{``If'' part}. As agent $i$ has at least two actions at $\boldsymbol{h}^*_i$, we can pick distinct $\theta^1_i,\theta^2_i \in \Theta_i(\boldsymbol{h}^*_i)$. Take $(\theta^1_j,\theta_{-i,j}) \in \Theta_{-i}(\boldsymbol{h}^*_i)$ and $(\theta^2_j,\theta_{-i,j}) \in (\Theta_{-i} \setminus \Theta_{-i}(\boldsymbol{h}^*_i))$ such that, without loss of generality, $f(\theta^1_j,\theta^1_i,\theta_{-i,j})P(\theta^{2}_j)f(\theta^{2}_j,\theta^{2}_i,\theta_{-i,j})$. Take a strategy $s_i \in S_i$ such that for any information set $\boldsymbol{h}_i$ of agent $i$, if $\boldsymbol{h}_i \preceq \boldsymbol{h}^*_i$ or $\boldsymbol{h}^*_i \prec \boldsymbol{h}_i$ with $\theta^1_i \in \Theta_i(\boldsymbol{h}_i)$, $\theta^1_i \in s_i(\boldsymbol{h}_i)$; otherwise, $\theta^2_i \in s_i(\boldsymbol{h}_i)$ whenever $\theta^2_i \in \Theta_i(\boldsymbol{h}_i)$. Let $s_{\theta^1_j}$, $s_{\theta^2_j}$, and $s_{\theta_{-i,j}}=(s_{\theta_k})_{k \neq i,j}$ be unconditional strategies for corresponding types. Then, we have $$\mathcal{X}(s_{\theta^1_j},s_i,s_{\theta_{-i,j}})=f(\theta^1_j,\theta^1_i,\theta_{-i,j})P(\theta^{2}_j)f(\theta^{2}_j,\theta^{2}_i,\theta_{-i,j})=\mathcal{X}(s_{\theta^2_j},s_i,s_{\theta_{-i,j}}),$$ which implies that $G$ is not incentive compatible. 

\textit{Contrapositive of ``only if''}. Suppose there exist $i \in N$, distinct $\theta^1_i,\theta^2_i \in \Theta_i$, and non-trivial partition $\{R^1_{-i},R^2_{-i}\}$ of $\Theta_{-i}$ such that for any $j \in N$ with $j \neq i$, any $(\theta^1_j,\theta_{-i,j}) \in R^1_{-i}$, any $(\theta^2_j,\theta_{-i,j}) \in R^2_{-i}$, and any $k,l \in \{1,2\}$, it is true that $f(\theta^k_j,\theta^l_i,\theta_{-i,j})R(\theta^{k}_j)f(\theta^{3-k}_j,\theta^{3-l}_i,\theta_{-i,j})$. We construct a non-static gradual mechanism $G$ in two steps. First, starting from the direct mechanism implementing $f$, apply an inverse COA with respect to agent $i$, such that (i) at the initial history, her action set becomes $A_i(\varnothing)=\big(\Theta_i \cup \{\{\theta^1_i,\theta^2_i\}\}\big)\setminus\{\theta^1_i,\theta^2_i\}$; and (ii)
if she chooses $\{\theta^1_i,\theta^2_i\}$, she subsequently selects between $\theta^1_i$ and $\theta^2_i$ at an information set $\boldsymbol{h}_i$ with $\Theta_{-i}(\boldsymbol{h}_i)=\Theta_{-i}$. Second, apply an ILL that partitions $\boldsymbol{h}_i$ into $\boldsymbol{h}^1_i$ and $\boldsymbol{h}^2_i$ with $\Theta_{-i}(\boldsymbol{h}^1_i)=R^1_{-i}$ and $\Theta_{-i}(\boldsymbol{h}^2_i)=R^2_{-i}$. Notice that in the intermediate gradual mechanism after the first step, for any $j \in N$ with $j \neq i$, only type profiles $\theta^1, \theta^2 \in \Theta$ with $\theta^1_{-i,j}=\theta^2_{-i,j}$ can be consistent with a common strategy profile $s_{-i,j}$ of agents other than $i$ and $j$. Therefore, the ILL in the second step is incentive-preserving, and $G$ is incentive compatible by Theorem \ref{theotrans}.
\end{proof}

The following Lemma \ref{lemmat2} is for the proof of Theorem \ref{theorp}. For any $\varnothing \neq M \subsetneq N$ and any history $h$, define $\Theta_{-M}(h)=\prod_{j \in N \setminus M}\Theta_j(h)$.

\begin{lemma}
\label{lemmat2}
For any two terminal histories $z^1, z^2\in Z$, if $z^1=z(s^1_M,s_{-M})$, $z^2=z(s^2_M,s_{-M})$, and $\Theta_{-M}(z^1) \cap \Theta_{-M}(z^2)=\varnothing$ for some $\varnothing \neq M \subsetneq N$, $s^1_M,s^2_M \in S_M$, and $s_{-M} \in S_{-M}=\prod_{j \in N \setminus M}S_j$, then there exist agent $j\in N\backslash M$ and two of her information sets $\boldsymbol{h}^1_j$ and $\boldsymbol{h}^2_j$ such that $\boldsymbol{h}^1_j\prec z^1$, $\boldsymbol{h}^2_j\prec z^2$, $\Theta_j(\boldsymbol{h}^1_j)=\Theta_j(\boldsymbol{h}^2_j)$, and $\boldsymbol{h}^1_j$ and $\boldsymbol{h}^2_j$ have the same immediate predecessor.
\end{lemma}

\begin{proof}

Let $z^1, z^2\in Z$ be two terminal histories such that $z^1=z(s^1_M,s_{-M})$, $z^2=z(s^2_M,s_{-M})$, and $\Theta_{-M}(z^1) \cap \Theta_{-M}(z^2)=\varnothing$ for some $s^1_M,s^2_M \in S_M$ and $s_{-M} \in S_{-M}$. For each $j \in N\backslash M$, define $\boldsymbol{H}^k_j=\{\boldsymbol{h}_j \in \boldsymbol{H}_j: \boldsymbol{h}_j \prec z^k\}$ for both $k=1,2$. Notice that $\{\varnothing\} \in \boldsymbol{H}^1_j \cap \boldsymbol{H}^2_j$. As $\Theta_{-M}(z^1) \cap \Theta_{-M}(z^2)=\varnothing$, there must be an agent $j^* \in N \setminus M$ such that $\Theta_{j^*}(z^1) \cap \Theta_{j^*}(z^2) =\varnothing$, which implies the existence of $\boldsymbol{h}^1_{j^*} \in \boldsymbol{H}^1_{j^*}$ and $\boldsymbol{h}^2_{j^*} \in \boldsymbol{H}^2_{j^*}$ such that $s_{j^*}(\boldsymbol{h}^1_{j^*}) \cap s_{j^*}(\boldsymbol{h}^2_{j^*}) =\varnothing$. Let $\bar{\boldsymbol{h}}_{j^*}$ be the $\preceq$-maximal element of $\boldsymbol{H}^1_{j^*} \cap \boldsymbol{H}^2_{j^*}$. For any predecessors $\boldsymbol{h}_{j^*}$ and $\boldsymbol{h}'_{j^*}$ of $\bar{\boldsymbol{h}}_{j^*}$, either $s_{j^*}(\boldsymbol{h}_{j^*}) \subseteq s_{j^*}(\boldsymbol{h}'_{j^*})$ or $s_{j^*}(\boldsymbol{h}'_{j^*}) \subseteq s_{j^*}(\boldsymbol{h}_{j^*})$. Therefore, there must exist two immediate successors of $\bar{\boldsymbol{h}}_{j^*}$ in $\boldsymbol{H}^1_{j^*}$ and $\boldsymbol{H}^2_{j^*}$, respectively, as predicated by the lemma.
\end{proof}

\begin{proof}[Proof of Theorem \ref{theorp}]
Let $f$ be a strategy-proof SCF and $G$ a gradual mechanism implementing $f$. We offer the following three observations. The first follows strategy-proofness of $f$, and the second and third rephrase reaction-proofness and Proposition \ref{Unconditional2}, respectively.
\begin{enumerate}
  \item for any two terminal histories $z^1$ and $z^2$ of $G$, if $\mathcal{X}(z^1) P(\theta_i) \mathcal{X}(z^2)$ for some agent $i$ and some $\theta\in \Theta(z^2)$, then $\Theta_{-i}(z^1)\cap \Theta_{-i}(z^2) = \varnothing$.

  \item $G$ is reaction-proof if and only if for any two terminal histories $z^1$ and $z^2$ such that (i) they are consistent with a common $s_{-i,j} \in S_{-i,j}$ and (ii) there exist $\boldsymbol{h}_i^1$ and $\boldsymbol{h}_i^2$ such that (a) $\boldsymbol{h}_i^1\prec z^1$, (b) $\boldsymbol{h}_i^2\prec z^2$, and (c) $\boldsymbol{h}_i^1$ and $\boldsymbol{h}_i^2$ have the same immediate predecessor and $\Theta_i(\boldsymbol{h}_i^1) = \Theta_i(\boldsymbol{h}_i^2)$, it is the case that $\mathcal{X}(z^1) R(\theta_j) \mathcal{X}(z^2)$ for any $\theta\in \Theta(z^1)$.
  
  \item $G$ is incentive compatible if and only if for any two terminal histories $z^1$ and $z^2$ consistent with a common $s_{-i} \in S_{-i}$, it is the case that $\mathcal{X}(z^1) R(\theta_i) \mathcal{X}(z^2)$ for any $\theta\in \Theta(z^1)$.
\end{enumerate}

If $G$ is reaction-proof, by Lemma \ref{lemmat2} and the first two observations, any pair of terminal histories $z^1$ and $z^2$ such that $\mathcal{X}(z^1) P(\theta_i) \mathcal{X}(z^2)$ for some agent $i$ and some $\theta\in \Theta(z^2)$ cannot be consistent with a common $s_{-i} \in S_{-i}$. Therefore, by the third observation, $G$ is incentive compatible. 

If $G$ is not reaction-proof, there exist two agents $i, j\in N$, a pair of agent $i$'s information sets $\boldsymbol{h}_i^1$ and $\boldsymbol{h}_i^2$ that have a common immediate predecessor with $\Theta_i(\boldsymbol{h}_i^1) = \Theta_i(\boldsymbol{h}_i^2)$, and a pair of type profiles $\theta^1\in \Theta(\boldsymbol{h}_i^1)$ and $\theta^2\in \Theta(\boldsymbol{h}_i^2)$ consistent with a common $s_{-i,j} \in S_{-i,j}$ such that $f(\theta^1) P(\theta_j^2) f(\theta^2)$. We can identify a strategy $s_i \in S_i$ that behaves as $s_{\theta^1_i}$ at any information set $\boldsymbol{h}_i$ with $\boldsymbol{h}^1_i \preceq \boldsymbol{h}_i$ and otherwise behaves as $s_{\theta^2_i}$, where $s_{\theta^1_i}$ and $s_{\theta^2_i}$ are unconditional strategies for $\theta^1_i$ and $\theta^2_i$, respectively. Take unconditional strategies $s_{\theta_j^1}$ and $s_{\theta_j^2}$ for $\theta^1_j$ and $\theta^2_j$, respectively. As $f(\theta^1)=\mathcal{X}(s_{\theta_j^1},s_i,s_{-i,j}) P(\theta_j^2) \mathcal{X}(s_{\theta_j^2},s_i,s_{-i,j})=f(\theta^2)$, $G$ is not incentive compatible.
\end{proof}

\begin{proof}[Proof of Theorem \ref{theo:IRP}]
  Arbitrarily take two agents $i$ and $j$, a pair of agent $i$'s information sets $\boldsymbol{h}_i^1$ and $\boldsymbol{h}_i^2$ that have a common immediate predecessor with $\Theta_i(\boldsymbol{h}_i^1) = \Theta_i(\boldsymbol{h}_i^2)$, a pair of histories $h^1 \in \boldsymbol{h}^1_i$ and $h^2 \in \boldsymbol{h}^2_i$ consistent with a common strategy profile $s_{-i,j} \in S_{-i,j}$, and a pair of type profiles $\theta^1\in \Theta(h^1)$ and $\theta^2\in \Theta(h^2)$ that are also consistent with $s_{-i,j}$. 

  Denote $z^1$ and $z^2$ the terminal histories such that $\theta^1 \in \Theta(z^1)$ and $\theta^2 \in \Theta(z^2)$. Let $M\subseteq N\backslash \{j\}$ be the collection of agent $k$ who has two distinct information sets $\boldsymbol{h}_k^1$ and $\boldsymbol{h}_k^2$ such that $\boldsymbol{h}_k^1\prec z^1$, $\boldsymbol{h}_k^2\prec z^2$, $\Theta_k(\boldsymbol{h}_k^1) = \Theta_k(\boldsymbol{h}_k^2)$, and $\boldsymbol{h}_k^1$ and $\boldsymbol{h}_k^2$ have the same immediate predecessor (notice that $i \in M$). Let $L = N\backslash (M \cup \{j\})$. For any pair of histories $h^1\preceq z^1$ and $h^2 \preceq z^2$, observe that (i) for each $k \in M$, $\Theta_k(h^1) \cap \Theta_k(h^2)=\varnothing$ only if $\boldsymbol{h}_k^1 \prec h^1$ and $\boldsymbol{h}_k^2 \prec h^2$; and (ii) for each $k \in L$, $\Theta_k(h^1) \cap \Theta_k(h^2) \neq \varnothing$.
  
  Note that, for both $l = 1, 2$, $f(\theta)R(\theta^*_j)f(\theta')$ for any $\theta,\theta' \in \Theta(z^l)$ and any $\theta^*_j \in \Theta_j$. Let $\underline{h}^1\preceq z^1$ and $\underline{h}^2\preceq z^2$ be the $\preceq$-minimal histories such that it is still the case that $f(\theta)R(\theta^*_j)f(\theta')$ for any $\theta,\theta' \in \Theta(\underline{h}^l)$ and any $\theta^*_j \in \Theta_j$, respectively for each $l\in \{1, 2\}$. 
  
  For any agent $k\in M$, let $h^1_k \in \boldsymbol{h}_k^1$ and $h^2_k \in \boldsymbol{h}_k^2$ be the histories such that $h^1_k \prec z^1$ and $h^2_k \prec z^2$. Take the strategy $s^*_i \in S_i$ that behaves as $s_{\theta^1_i}$ at any information set $\boldsymbol{h}_i$ with $\boldsymbol{h}^1_i \preceq \boldsymbol{h}_i$ and otherwise behaves as $s_{\theta^2_i}$. Notice that $h_k^1$ and $h_k^2$ are consistent with the strategy profile $(s^*_i,s_{-i,j,k})$. Suppose $G$ is strongly reaction-proof. Then, there is no $k\in M$ such that $\boldsymbol{h}_k^1\prec \underline{h}^1$ and $\boldsymbol{h}_k^2\prec \underline{h}^2$, otherwise for some $l \in \{1,2\}$, $\underline{h}^l$ cannot be the $\preceq$-minimal history such that $f(\theta)R(\theta^*_j)f(\theta')$ for any $\theta,\theta' \in \Theta(\underline{h}^l)$ and any $\theta^*_j \in \Theta_j$. By the previous observations, we have $\Theta_{-j}(\underline{h}^1)\cap \Theta_{-j}(\underline{h}^2)\neq \varnothing$. For any $\theta_{-j}^*\in \Theta_{-j}(\underline{h}^1)\cap \Theta_{-j}(\underline{h}^2)$, by strategy-proofness of $f$, we have 
  \begin{equation*}
    \underbrace{f(\theta^1) R(\theta_j^1)f(\theta_j^1, \theta_{-j}^*)}_{\theta^1, (\theta^1_j, \theta^*_{-j}) \in \Theta(\underline{h}^1)} R(\theta_j^1)\underbrace{f(\theta_j^2, \theta_{-j}^*) R(\theta_j^1) f(\theta^2)}_{\theta^2, (\theta^2_j, \theta^*_{-j}) \in \Theta(\underline{h}^2)}.
  \end{equation*}
  Thus, strong reaction-proofness implies reaction-proofness. 
\end{proof}

\subsection{Proofs for Section \ref{secapp}}

In this subsection, we will use the following construction and observation. Let $\boldsymbol{h}^1_i$ and $\boldsymbol{h}^2_i$ be two information sets of agent $i$ that have a common immediate predecessor with $\Theta_i(\boldsymbol{h}^1_i)=\Theta_i(\boldsymbol{h}^2_i)$. Let $h^1\in \boldsymbol{h}_i^1$ and $h^2 \in \boldsymbol{h}_i^2$ be consistent with a common strategy profile $s_{-i,j} \in S_{-i,j}$. Define $H^1=\{h \in H: h \preceq h^1\}$ and $H^2=\{h \in H: h \preceq h^2\}$. Let $\underline{h}$ be the $\preceq$-maximal element in $H^1 \cap H^2$, and $\underline{h}^1$ and $\underline{h}^2$ be the immediate successors of $\underline{h}$ in $H^1$ and $H^2$, respectively. Then, it must be the case that $\Theta_j(\underline{h}^1) \neq \Theta_j(\underline{h}^2)$, i.e., $j$'s different actions at $\underline{h}$ lead to the first divergence of the paths from the initial history to $h^1$ and $h^2$.

\label{proofsec4}
\begin{proof}[Proof of Proposition \ref{propapaic}.]
  We first show the incentive compatibility of the open-by-two ascending-price auction. Let $\boldsymbol{h}^1_i$ and $\boldsymbol{h}^2_i$ be two information sets of bidder $i$ that have a common immediate predecessor where bidder $i$ chose to stay in the auction at the price level $p - 1$. Suppose $\boldsymbol{h}^1_i$ and $\boldsymbol{h}^2_i$ contain a pair of histories $h^1$ and $h^2$, respectively, consistent with a common $s_{-i,j} \in S_{-i,j}$. As the first divergence of $h^1$ and $h^2$ results from bidder $j$'s different actions at a common predecessor $\underline{h}$, bidder $j$ has left the auction before one of the histories. Without loss of generality, assume that bidder $j$ leaving the auction at a price level $p'$ is a part of the history $h^1$. Define $\underline{H}^1=\{h \in H: \underline{h} \prec h \preceq h^1\}$ and $\underline{H}^2=\{h \in H: \underline{h} \prec h \preceq h^2\}$.
  \begin{enumerate}
    \item If $p'<p$, it is obvious that $j$ has lost the auction at $h^1$.
    \item If $p'=p$, it is also the case that $j$ has lost the auction at $h^1$ since there must be another bidder $k$ with $k < i$ who has chosen to stay at $p$ before both $h^1$ and $h^2$. Otherwise, there is a bijection $m: \underline{H}^1\rightarrow \underline{H}^2$ respecting the precedence relation on histories such that for any $h \in \underline{H}^1$, $h$ and $m(h)$ are in the same information set of some bidder. This contradicts that $h^1$ and $h^2$ belong to different information sets of bidder $i$.

  \end{enumerate}

Note that $\boldsymbol{h}^1_i$ and $\boldsymbol{h}^2_i$ and the corresponding $h^1$ and $h^2$ are chosen arbitrarily. Therefore, we have shown that the open-by-two ascending-price auction is strongly reaction-proof. By Theorem \ref{theo:IRP}, it is incentive compatible.

Next, we prove that the open-by-two ascending-price auction is transparent. Let $\boldsymbol{h}_i$ be a non-singleton information set of bidder $i$ associated with a price level $p<m$. Let $h^*\in \boldsymbol{h}_i$ be the history where all bidders indexed before $i$ have left the auction at $p$. Let $\boldsymbol{h}_i^1$ and $\boldsymbol{h}_i^2$ be two non-empty disjoint subsets of $\boldsymbol{h}_i$ with their union being $\boldsymbol{h}_i$. Assume, without loss of generality, that $h^* \in \boldsymbol{h}_i^1$. Take a history $h'\in \boldsymbol{h}_i^2$. Then there exists a unique bidder $j$ (with $j < i$) who has decided to stay in the auction at the price level $p$.

 Let $v_i^1=v_j^1=p$ and $v_i^2=v_j^2=p+1$. There exists $v_{-i,j} \in V^{n-2}$ with $v_k \leq p$ for each $k\neq i, j$ such that $v^1=(v_i^1, v_j^1, v_{-i,j})\in \Theta(h^*) \subseteq \Theta(\boldsymbol{h}_i^1)$ and $v^2=(v_i^2, v_j^2, v_{-i,j})\in \Theta(h') \subseteq \Theta(\boldsymbol{h}_i^2)$. Both $v^1$ and $v^2$ are consistent with $(s_{v_k})_{k \in N \setminus \{i,j\}}$, where $s_{v_k}$ is an unconditional strategy for bidder $k$'s type $v_k$. However, $EV(v^1| v_j^2)=1/|k \in N: v_k^1 = p| > 0 = EV(v^2|v^2_j)$, meaning that the ILL splitting $\boldsymbol{h}_i$ into $\boldsymbol{h}_i^1$ and $\boldsymbol{h}_i^2$ is not incentive-preserving.
    \end{proof}

The following lemmas \ref{lemmardaimplementher} and \ref{lemmardasrp} are for the proof of Proposition \ref{proprda}.

    \begin{lemma}
      Given an HER $f^{\Gamma}$ identified by inheritance structure $\Gamma$ and an RDA mechanism for $f^{\Gamma}$, $f^{\Gamma}(\theta)=\mathcal{X}(s_{\theta})$ for any $\theta \in \Theta$, where $s_{\theta}=(s_{\theta_i})_{i \in N}$ is a profile of unconditional strategies for the corresponding types.
      \label{lemmardaimplementher}
  \end{lemma}
  
  \begin{proof}
      Let $\Gamma$ be an inheritance structure, $f^{\Gamma}$ an HER, and $G$ an RDA mechanism for $f^{\Gamma}$. In this proof, we write a trading cycle as $\{j_1,o_1,...,j_g,o_g\}$ where $j_s$ receives the object $o_s$ owned by $j_{s+1}$ (i.e., $o_s$ is in the endowed menu of $j_{s+1}$) for each $s=1,...,g$, with $j_{g+1}=j_1$. We show that, for any $\theta \in \Theta$ and any trading cycle $C$ formed in the algorithm of $f^{\Gamma}$ with input $\theta$, the same cycle $C$ will form if agents play $s_{\theta}$ in $G$.
  
  
      For any stage $t \geq 1$, let $C_{t}=\{j_1,o_1,...,j_g,o_g\}$ be a trading cycle formed in stage $t$ (in the algorithm of $f^{\Gamma}$ with input $\theta$) and $\mathcal{C}^{t-1}$ be the collection of all the trading cycles formed in stages $1,...,t-1$ (let $\mathcal{C}^{0}=\varnothing$). Let $t^G$ be some stage (along the path in $G$ determined by $s_{\theta}$) such that all the trading cycles in $\mathcal{C}^{t-1}$ have formed and have been removed from the market. Note that $t^G$ exists for $t=1$, and the following induction guarantees the existence of $t^G$ for any stage $t$. Denote $x$ and $x^G$ the partial allocations realized before stage $t$ and before stage $t^G$, respectively.
  
      \medskip
  
      \noindent \textit{Claim $1$.} In any stage $t'^{G}\leq t^G$, for any object $o_{s} \in C_t$, if agent $j_{s+1} \in C_t$ remains in the market, then $o_s$ also remains and is owned by either $j_{s+1}$ or some agent in $\bigcup \mathcal{C}^{t-1}$. 
      
      Since $o_s \in \mathcal{E}_{j_{s+1}}(x)$, either $o_s \in \mathcal{E}_{j_{s+1}}(\varnothing)$, or there is a path in $\Gamma_{o_s}$, denoted by a sequence of agents $(k_1,...,k_g,j_{s+1})$ with $\{k_1,...,k_g\} \subseteq \bigcup \mathcal{C}^{t-1}$, such that $o_s$ is initially owned by $k_1$, and for each $k_r \in \{k_1,...,k_g\}$, the next agent inherits $o_s$ since $k_r$ is assigned to $x(k_r)$. Claim $1$ follows that $x^{G}$ is an extension of $x$. 
  
      \medskip
  
      \noindent \textit{Claim $2$.} For any agent $j_s \in C_{t}$ and object $o_s \in C_t$, $o_s P(\theta_{j_s}) o$ for any $o \in O \setminus \bigcup\mathcal{C}^{t-1}$ with $o \neq o_s$.
  
      Claim $2$ directly follows that in stage $t$, (i) the set of remaining objects is $O \setminus \bigcup\mathcal{C}^{t-1}$, and (ii) for each agent $j_s$ in $C_t$, she receives her most preferred remaining object.
  
      \medskip
  
      There are two cases.
  
      \noindent \textit{Case $1$.} \textit{all agents in $C_t$ remain at the beginning of stage $t^G$}. For any $s=1,...,g$, by Claim $1$, $o_s$ is owned by $j_{s+1}$ in stage $t^G$. By Claim $2$, under $s_{\theta_{j_s}}$, if $j_s$ is a standby owner in stage $t^G$, she will always choose the menu containing $o_s$ whenever $o_s$ remains in the market; if she is not involved, she must have designated the owner of $o_s$ as trading partner. Therefore, $C_t$ will form in stage $t^G$ or later.
  
      \medskip
      \noindent \textit{Case $2$.} \textit{some agent in $C_t$ has left the market before stage $t^G$}. Let $t'^{G}<t^G$ be the ealierst stage such that a trading cycle $C_{t'^{G}}$ containing some agent in $C_t$ forms. Let $j_s$ be any agent in $C_t \cap C_{t'^{G}}$ and $o_{s'}$ be the object assigned to $j_s$ in $C_{t'^{G}}$. 
      
      First, we show $o_{s'}=o_s$. Since agent $j_{s+1} \in C^t$ remains in stage $t'^{G}$, by Claim $1$, object $o_s \in C^t$ must remain in stage $t'^{G}$. Then, by Claim $2$ and the fact that every agent $i$ plays $s_{\theta_i}$, we have $o_{s'} \neq o$ for any object $o \in O \setminus \bigcup \mathcal{C}^{t-1}$ with $o \neq o_s$. Furthermore, as $j_s \notin \bigcup \mathcal{C}^{t-1}$, $o_{s'} \neq o$ for any object $o \in \bigcup \mathcal{C}^{t-1}$. Therefore, $o_{s'}=o_s$.
  
      Second, we show that any agent in $C_t$ must also be in $C_{t'^{G}}$. Suppose not, then there must be agents $j_r$ and $j_{r+1}$ in $C_t$ with $j_r \in C_{t'^{G}}$ and $j_{r+1} \notin C_{t'^{G}}$. By the first step and Claim $1$, some agent $i \in \bigcup \mathcal{C}^{t-1}$ must be in $C_{t'^{G}}$ as the owner of $o_r$. However, $j_r \notin \bigcup \mathcal{C}^{t-1}$ implies $C_{t'^{G}} \notin \mathcal{C}^{t-1}$, which means that agent $i$ appears in two distinct trading cycles formed along the path in $G$ determined by $s_{\theta}$, a contradiction.
  
      By the first two steps, each agent $j_s \in C_t$ belongs to $C_{t'^{G}}$ and receives the same object $o_s$ in both $C_t$ and $C_{t'^{G}}$. To see $C_{t'^{G}}=C_t$, it suffices to show that in stage $t'^{G}$, any object $o_s \in C_t$ is owned by agent $j_{s+1} \in C_t$, which is implied by Claim $1$ and the fact that no agent in $\mathcal{C}^{t-1}$ is in $C_{t'^{G}}$.
  \end{proof}

\begin{lemma}
  The RDA mechanism for any HER is strongly reaction-proof.
  \label{lemmardasrp}
\end{lemma}
 
  \begin{proof}

  In an RDA mechanism $G$ for some HER, let $\boldsymbol{h}^1_i$ and $\boldsymbol{h}^2_i$ be two information sets of some agent $i$ with $\Theta_i(\boldsymbol{h}^1_i)=\Theta_i(\boldsymbol{h}^2_i)$ that share a common immediate predecessor. Let $h^1 \in \boldsymbol{h}^1_i$ and $h^2 \in \boldsymbol{h}^2_i$ be histories consistent with some $s_{-i,j} \in S_{-i,j}$ for some agent $j \neq i$. We show the strong reaction-proofness of $G$ by proving that $j$ must have left the market with an assigned object before either $h^1$ or $h^2$.

  Suppose, toward a contradiction, that $j$ remains in the market at both
  $h^1$ and $h^2$. For $l=1,2$, let $H^l=\{h\in H: h\preceq h^l\}$,
  and let $H^{l,t}$ and $H^{l,<t}$ denote, respectively, the histories in $H^l$ that lie in stage $t$ and strictly before stage $t$; whenever $H^{l,t}\neq \varnothing$, let $\underline h^{l,t}$ be the earliest history in $H^{l,t}$; and denote $C^{l,t}$ the trading cycle formed in stage $t$ along the
  path to $h^l$ if that cycle has formed weakly before $h^l$.

For $\hat{H}^1 \subseteq H^1$ and $\hat{H}^2 \subseteq H^2$ with equal cardinality, a bijection $m:\hat{H}^1 \to \hat{H}^2$ is an \emph{order-preserving bijection} if for every $h \in \hat{H}^1$:  
(i) the sets of active agents at $h$ and $m(h)$ coincide;  
(ii) for each active agent $k$, $h$ and $m(h)$ belong to the same information set $\boldsymbol{h}_k$ at which $k$ takes the same action along the two paths;  
(iii) $h \prec \bar{h}$ implies $m(h) \prec m(\bar{h})$.

We proceed by induction on stages. Fix $t \ge 1$ such that (i) $j$ remains in the market at both $\underline{h}^{1,t}$ and $\underline{h}^{2,t}$; (ii) when $t>1$, $h^1 \notin H^{1,<t}$ and $h^2 \notin H^{2,<t}$; (iii) there exists an order-preserving bijection $m^{<t}:(H^{1,<t} \setminus H_j) \to (H^{2,<t} \setminus H_j)$ whenever $(H^{1,<t} \setminus H_j) \neq \varnothing$, where $H_j$ is the set of histories at which $j$ is active. The induction hypothesis implies that, at the beginning of stage $t$ along the paths to $h^1$ and $h^2$, remaining agents and their endowed menus, standby owners excluding $j$, and previous choices by agents other than $j$ are identical. We show that $h^1 \notin H^{1,t}$ and $h^2 \notin H^{2,t}$.

\medskip

\noindent
\textit{Case 1.} $h^1 \in H^{1,t}$ and $h^2 \in H^{2,t}$.  
Since $\boldsymbol{h}^1_i$ and $\boldsymbol{h}^2_i$ follow the same action at a common immediate predecessor, by implications of the induction hypothesis, $h^1$ and $h^2$ must be at the same sub-stage and with identical available actions. By the information structure of the RDA mechanism, $h^1$ and $h^2$ belong to the same information set of $i$, contradicting $\boldsymbol{h}^1_i \neq \boldsymbol{h}^2_i$.

\medskip

\noindent
\textit{Case 2.} $h^1 \in H^{1,t}$ and $h^2 \in H^{2,t'}$ for some $t'>t$.  

(i) If $j \in C^{2,t}$, then $j$ has left the market before $h^2$, a contradiction.

(ii) If $j \in C^{1,t}$ but $j \notin C^{2,t}$, let $h^{1,t} \in H^{1,t}$ be the history at which the active agent's action closes the trading cycle $C^{1,t}$. Note that $h^{1,t} \neq \underline{h}^{1,t}$, otherwise $C^{1,t}=\{j\}$ and no agent other than $j$ is active at any history in $H^{1,t}$. 

Consider the sequence of histories $\{h \in (H^{1,t} \setminus H_j) : \underline{h}^{1,t} \preceq h \preceq h^{1,t}\}$, which contains at least $\underline{h}^{1,t}$. As standby owners move sequentially in index order within the renunciation and designation sub-stages, the $l$-th history in the sequence corresponds to the $l$-th history in $H^{2,t} \setminus H_j$ where the same active agent $k$ chooses from the same set of actions; if such a counterpart does not exist, the only reason is that a trading cycle containing $j$ forms at some history in $H^{2,t}$, contradicting the present case.

By the information structure, the two histories in each corresponding pair belong to the same information set $\boldsymbol{h}_k$ of some agent $k\neq j$. By the common strategy profile $s_{-i,j}$ for $k\neq i,j$, and by the construction of $\boldsymbol{h}_i^1$ and $\boldsymbol{h}_i^2$ for $k=i$, agent $k$ must take the same action at $\boldsymbol{h}_k$ along the paths to both $h^1$ and $h^2$. Therefore, there is an order-preserving bijection between $\{h \in (H^{1,t} \setminus H_j): \underline{h}^{1,t} \preceq h \preceq h^{1,t}\}$ and the first 
$\left| \{h \in (H^{1,t} \setminus H_j) : \underline{h}^{1,t} \preceq h \preceq h^{1,t}\} \right|$ histories in $H^{2,t} \setminus H_j$. It follows that $h^1$ must be in the assertion sub-stage; otherwise $h^1$ and a predecessor of $h^2$ would be in the same information set, contradicting that $\boldsymbol{h}^1_i$ and $\boldsymbol{h}^2_i$ have a common immediate predecessor.

Write $C^{1,t}=\{k_1,\dots,k_g\}$ with $k_g=j$. By the order-preserving bijection defined above and $m^{<t}$, each $k_r \neq j$ (including $i$) must either designate $k_{r+1}$ in stage $t$ or have designated $k_{r+1}$ before stage $t$ along the path to $h^2$. Hence $h^2$ is in an assertion sub-stage or in a renunciation sub-stage. In the former case, $h^1$ and $h^2$ are in the same information set; in the latter, $j$ has left the market before $h^2$. Both are contradictions.

(iii) If $j \notin (C^{1,t} \cup C^{2,t})$, the construction of order-preserving bijection in Case $2$(ii) applies in both directions, yielding an order-preserving bijection between $H^{1,t} \setminus H_j$ and $H^{2,t} \setminus H_j$. This implies that $h^1$ and a predecessor of $h^2$ lie in the same information set, thus $\boldsymbol{h}^1_i$ and $\boldsymbol{h}^2_i$ cannot have a common immediate predecessor.

\medskip

\noindent \textit{Case 3.} The case $h^2 \in H^{2,t}$ and $h^1 \in H^{1,t'}$ for $t'>t$ is symmetric to Case $2$.

\medskip

We have shown that $h^1 \notin H^{1,t}$ and $h^2 \notin H^{2,t}$, implying $j \notin (C^{1,t} \cup C^{2,t})$. By Case $2$(iii), there is an order-preserving bijection $m^{<t+1}: (H^{1,<t+1}\setminus H_j) \to (H^{2,<t+1}\setminus H_j)$. By induction, $h^1 \notin H^{1,t}$ and $h^2 \notin H^{2,t}$ for any stage $t$. As the mechanism is finite, we have $h^1 \notin H^1$ and $h^2 \notin H^2$, a contradiction.
  \end{proof}

\begin{proof}[Proof of Proposition \ref{proprda}]
  Proposition \ref{proprda} holds by lemmas \ref{lemmardaimplementher}, \ref{lemmardasrp}, and theorems \ref{theorp}, \ref{theo:IRP}.
\end{proof}
 
\clearpage
\begingroup

\setlength{\bibsep}{0pt}    

\setlength{\itemsep}{0pt}  

\renewcommand{\baselinestretch}{1.1}\selectfont

\bibliography{DTGMref}

\endgroup

\end{document}